\documentclass[galaxies,article,submit,moreauthors,pdftex]{Definitions/mdpi} 
\usepackage{amssymb}

\firstpage{1} 
\pubvolume{8}
\issuenum{3}
\articlenumber{66}
\pubyear{2020}
\copyrightyear{2020}
\history{Received: date; Accepted: date; Published: date}

\Title{Short-term X-ray Variability during Different Activity Phases of Blazars
S5 0716+714 and PKS 2155-304}

\Author{Pankaj Kushwaha $^{1,\dagger,*}$\orcidA{}, and Main Pal $^{2}$\orcidB{}}

\AuthorNames{Pankaj Kushwaha and Main Pal}

\address{%
$^{1}$ \quad Aryabhatta Research Institute of Observational Sciences (ARIES), Manora Peak, Nainital 263002, India; pankaj.kushwaha@aries.res.in\\
$^{2}$ \quad Centre for Theoretical Physics, Jamia Millia Islamia, New Delhi 110025, India; rajanmainpal@gmail.com\\
$^\dagger \quad $Aryabhatta Postdoctoral Fellow}

\corres{Correspondence: pankaj.kushwaha@aries.res.in}

\abstract{We explored the statistical properties of short-term X-ray variability
using long-exposure {\it XMM-Newton} data during high X-ray variability phases
of blazars S5 0716+714 and PKS 2155-304. In general,  hardness ratio
shows correlated variations with the source flux state (count rate), but in a
few cases, mainly the bright phases, the trend is
complex with correlation and anti-correlation both, indicating spectral evolution. 
Stationarity tests suggest the time series as non-stationarity or have trend 
stationarity. Except for one, none of the histograms fit resulted in a reduced-\(\chi^2
\sim 1\) for a normal and log-normal profile but a normal profile is favored in general.
On the contrary, the Anderson-Darling test favors lognormal with a test-statistic
value lower for log-normal over normal for all the
observations, even if out of significance limits. None of the IDs show linear RMS-flux
relation. The contrary inferences from widely used different statistical methods
indicate that a careful analysis is needed while the complex behavior of count
rate with hardness ratio suggests spectral evolution over a few 10s of
kilo-seconds during bright phases of the sources. In these cases, the spectrum 
extracted from whole observation may not be meaningful for spectral studies and
certainly not a true representation of the spectral state of the source.}


\begin{document}

\section{Introduction}
Strong and rapid variability has been one of the defining criteria of blazars --
active galactic nuclei with a powerful relativistic jet of plasma directed within
a few degrees with our line of sight. Studies focusing on temporal flux variability
have found them to be variable on all timescales accessible to us -- from the shortest
allowed by the observing facilities to the longest allowed by the archived data
\citep[e.g.][]{2018ApJ...863..175G}. Though the processes responsible for the observed
behavior are still unclear except for a general understanding, studies of these
observations have revealed some features that have resulted in empirical classification
schemes to classify and identify astrophysical objects. In the blazar parlance,
the observed variability on all timescales probed so far are now widely referred
into three categories: intra-day variability \citep[IDV;][]{1995ARA&A..33..163W} -- variability within a day, short-term
variability (STV) -- variability on days to months, and long-term variability (LTV)
-- variability over a few months and longer. Among these, IDV is now one of
the tools to identify an extra-galactic source as blazars. A spectacular prime
example of this in recent times has been in establishing a subset of narrow-line Seyfert galaxies (NLS) as blazars \citep{2007ApJ...658L..13Z,2008ApJ...685..801Y}.

Earlier and even now, optical bands have been primarily leading the IDV studies as
photon statistic at higher energies has prohibited such studies in the past. However,
with the continuous advancement in detector technologies, such sub-day studies
are now been possible at X-ray \citep[e.g.][]{2005ApJ...629..686Z,2005A&A...443..397B,
2010ApJ...718..279G,2020ApJ...897...25B},
gamma-ray \citep[e.g.] []{2018ApJ...854L..26S,2014ApJ...796...61K}, and very high
energies \citep[VHEs;][]{2007ApJ...664L..71A,2007ApJ...669..862A}, albeit still for
bright flares. These studies have also reported sub-hours to a few minutes variability
-- implying a compact, highly luminous emission region, still optically thin to pair
production. Combined with multi-wavelength (MW) observations on similar time scales,
these studies have been invaluable in providing insights on the nature of variability
and the underlying physical processes.

Variability reflects a change in the dynamical state of the system responsible for
the observable and thus, offers direct access to scales and dynamics of the system
except that in non-linear systems, a one-to-one relation between the observable
and underlying dynamics is not clear apparently \citep[e.g.][]{2020Galax...8...15K}.
Similar
problems exist for statistical exploration -- widely different systems with almost
no common physical connection can exhibit similar statistical behavior \citep[e.g.]
[]{2007HiA....14...41Z,2020ApJ...895...90S}. However, based on inferences/understanding from previous
studies, a combination of both can be used to explore specific aspects of different
problems that may offer new insights.

A major challenge in exploring blazars variability has been the enormous range 
of time scales involved and irregular gaps in the data which is inevitable
in astronomy.
Due to this, the statistical properties of blazars' MW emission have not been explored
widely until very recently. The renewed focus and targeted campaigns of blazars
with good cadence observations, especially after the launch of the {\it Fermi}
observatory that continuously surveys the sky at MeV-GeV energies have made such
studies now possible, over both long and short timescales \citep[e.g.][]{2014ApJ...786..143S,2018ApJ...863..175G}.
Attempts have been made to connect the short- and long-term behavior \citep[e.g.]
[]{2018ApJ...859L..21C,2020ApJ...897...25B} and compare these properties to those
of other AGNs. Though they can provide a general idea, inferring any time scale
 by comparing long- and short-term trends are fraught with uncertainties as
the short-term behavior is not yet widely established.
Different studies exploring short-term behaviors have reported significant changes
in these properties \citep{2005A&A...443..397B,2005ApJ...629..686Z} over different
time scales. Further, irregular gaps and statistical properties of time series
e.g. stationarity, limit the use of Fourier based methods and reliability of
inferences derived from it \citep[e.g.][]{2018ApJ...863..175G}.

The short-term variability being primarily a result of competition between particle
acceleration and cooling provides not only a key to study these two but the
statistical behavior can be used to test a few models of blazars as well a comparative 
study with the well established properties of AGNs. In this sense, long-term properties
of variability have been explored fairly well \citep[e.g.][]{2011A&A...531A.123K,2014ApJ...786..143S,
2016ApJ...822L..13K,2017ApJ...849..138K,2018ApJ...863..175G,2019ApJ...885...12R}
but the short-term detailed studies are still a few \citep[e.g.][]{2005A&A...443..397B,
2005ApJ...629..686Z,2013A&A...558A..92B,2018ApJ...864..164Y,2020ApJ...897...25B}, especially at high
energies due to fewer long exposure observations during such occurrences. Interestingly,
literature refers to blazars as extreme AGNs with emission dominated almost entirely
by the relativistic jet, in stark
contrast with a majority of AGNs. Yet in the terms of statistical properties of
variability, they appear similar to other AGNs -- stochastic variability with a
statistical trend similar to those exhibited by the accretion-powered sources in
general \citep[and references therein]{2016ApJ...822L..13K,2017ApJ...849..138K}.
This has ignited the study of disk-jet connection in blazars as a multiplicative
combination of fluctuations in the accretion-disk has been one of the widely accepted
explanations for such behavior in AGNs (and accretion-powered sources in general) \citep{1997MNRAS.292..679L,2005MNRAS.359..345U}. However, recently, based on simulations
and the fact that blazars jets are highly magnetized, additive model like minijets-in-a-jet
\citep{2012A&A...548A.123B} has been found to exhibit these characteristics. The
statistical behavior in this model changes drastically depending on the number of
regions contributing to the emission. Studies over short duration during high variability
phases can be used to test this model and also explore/establish general
behaviors.

In this work, we explore the statistical properties of X-ray variability over a
scale of a few minutes to days, during the highly variable episodes of blazar
S5 0716+714 and PKS 2155-304. The next section presents the details of data acquisition,
reduction procedures, and the selection of the sources followed by our
analyses and results in \S\ref{sec:Analysis}. We discuss our findings, its
implications, and the conclusion in section \S\ref{sec:Discuss}.

\section{{\it XMM-Newton} X-ray Data}\label{sec:reduction}
As our focus is investigations of statistical properties over short time scales
(a few minutes to hours), the 
observations chosen are biased in the sense that only long exposure observations
with good photon counts and exhibiting variability has been used. A high photon
count rate and long exposure allow binning that results in more data points needed
for statistical studies when the source has shown variability. Table \ref{tab:obslog} lists the sources
and the observations used in this work. All the observations are taken from the public
archive of the {\it XMM-Newton} observatory.

\begin{table}[H]
\caption{Details of {\it XMM-Newton} observations considered in this work}
\centering
\begin{tabular}{ccccccc}
\toprule
\textbf{Source}	& \textbf{Observation-ID}	& \textbf{Date} & Duration & GTIs$^*$ & Filter & rate (count s$^{-1}$)\\
\midrule
S5~0716+714	& 0502271401 & 2007-09-24 16:23:32 & 73917 & 50120&Thin1&5.0$\pm 0.6$ \\
PKS~2155--304& 0124930301 &2001-11-30 02:36:09& 92617 & 43465&Medium&62.2$\pm 0.1$\\
            & 0411780401 &2009-05-28 08:29:11 & 64820 & 41966 &Medium&61.2$\pm 0.1$\\
            & 0411780501 &2010-04-28 23:54:50& 74298 & 59670 &Medium&31.60$\pm 0.03$\\
            & 0124930601 &2002-11-29 23:27:28& 114675 & 56741&Thick&29.8$\pm 0.03$\\
            & 0411780701 &2012-04-28 00:48:26& 68735 & 47228&Medium&11.82$\pm 0.02$\\ 
            & 0411782101 &2013-04-23 22:31:38& 76015 & 61753&Medium&24.40$\pm 0.11$\\
            & 0727770901 &2014-04-25 03:14:56&65000 & 60163&Medium&29.40$\pm 0.03$\\
            & 0411780101 &2006-11-07 00:22:47& 101012 & 29369&Thin1&42.7$\pm 0.1$\\
            &  & & & 34745&Medium&45.3$\pm 0.1$\\
            &  & & & 35133&Thick&36.5$\pm 0.1$\\
\bottomrule
$^*$ seconds
\end{tabular}
\label{tab:obslog}
\end{table}

{\bf Reduction:} We used the {\it XMM-Newton} Science Analysis System (SAS) version
15.0.0 and followed the standard data reduction procedure for the light curve
extraction and analysis as described in the ``{\it XMM-Newton ABC Guide''}. Firstly,
for each ID, we generated a summary of the Observation Data File (ODF) and
calibration index file (CIF) using the latest calibration data files. We then used
the {\tt epproc} tool to reprocess the data from the European Photon Imaging Camera-PN
detector as it has a large collecting area in small window mode. We investigated
the background light curve in the 10-12 keV band to find intervals affected
by particle/solar flares. Subsequently, such intervals were removed from the data
to get good time intervals (GTIs). We selected single and double events to mitigate
the pile-up issue and we checked this problem by using the {\tt EPATPLOT} tool.
Except the ID 0124930301, none showed any significant pile-up. We removed a circular
core of radius 7 arcsec to minimize the pile-up events in this particular observation
ID. We then generated clean events files and the resulting GTIs are shown in Table
\ref{tab:obslog}. To extract the source light curve, we used a circular region
of radius between 40-45 arcsec centered at the source coordinates. For background,
we used circular regions with radius in the range of 40-45 arcsec, well away
from the source and any bright spot on the CCD.

To generate light curves, we used the HEASARC tool {\tt XSELECT} on the clean event
files and selected the source and background region. We derived light curves in
three energy bands: 0.3-2 keV, 2-10 keV, and 0.3-10 keV bands. The background-subtracted
source light curves were extracted using the ftool task {\tt lcmath}.
We also used the ftool task {\tt lcurve} to derive the hardness ratio (HR) between
the 0.3-2 keV and 2-10 keV energy bands.

\begin{figure}
\centering
\includegraphics[scale=0.29]{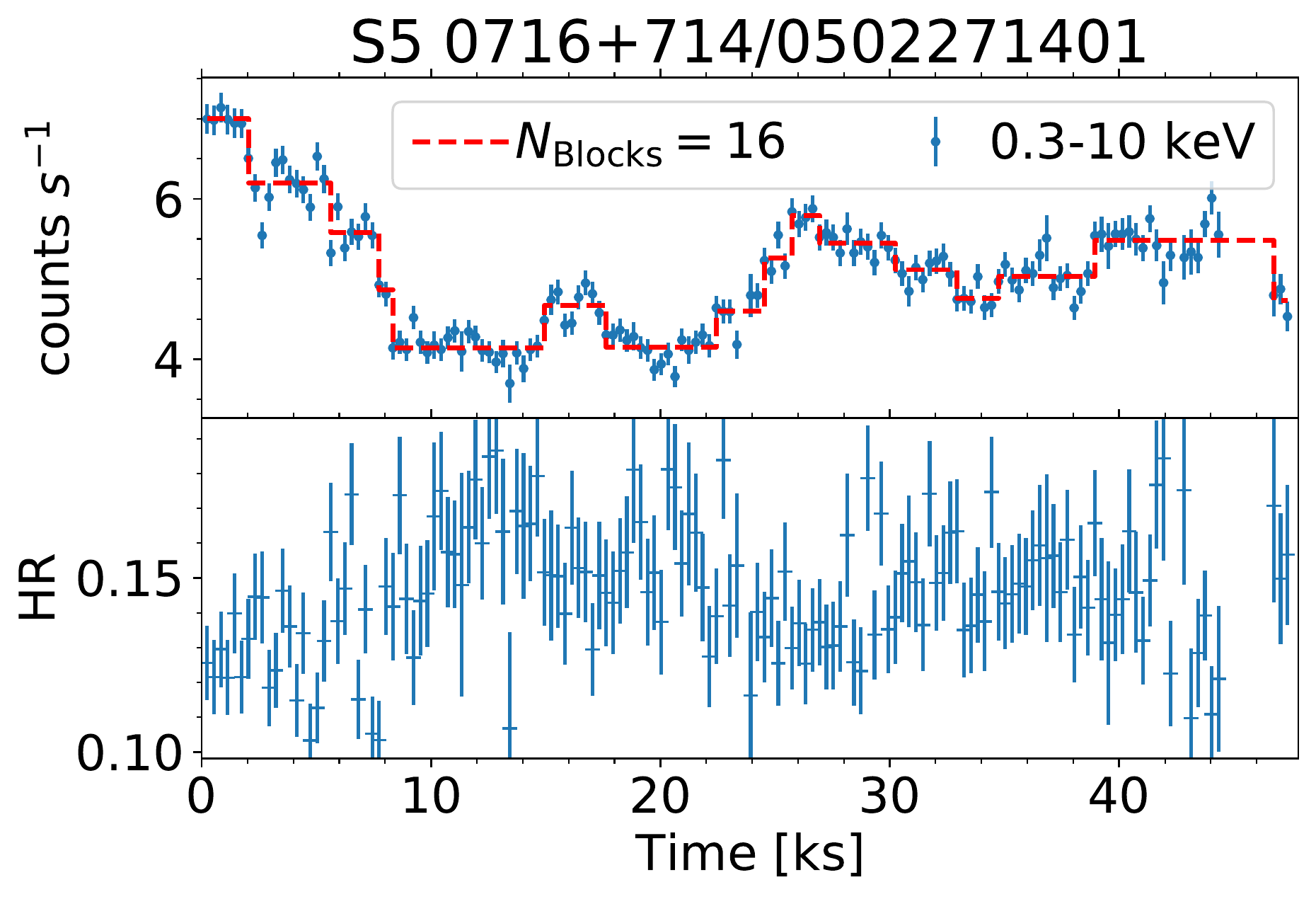}
\includegraphics[scale=0.29]{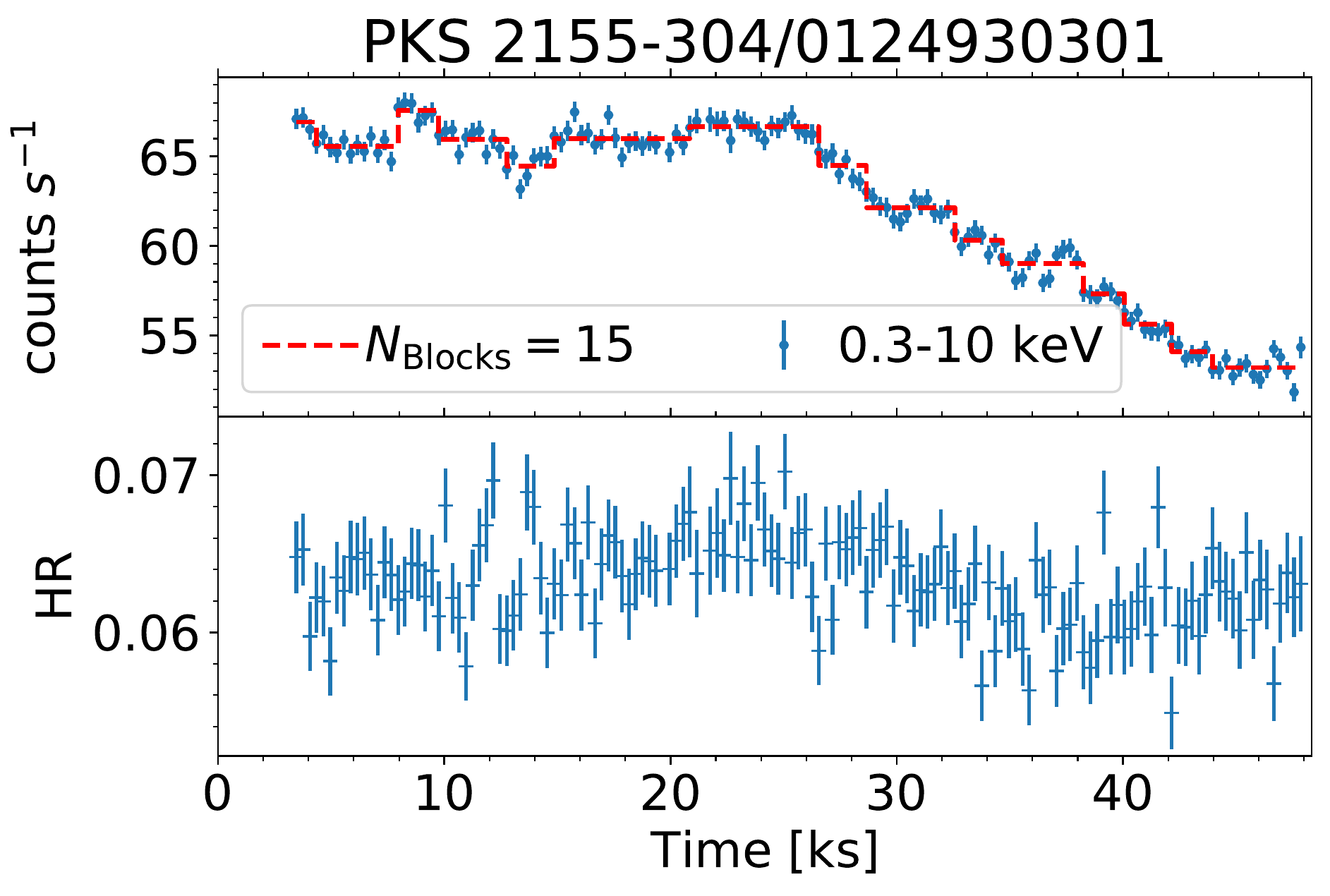}
\includegraphics[scale=0.29]{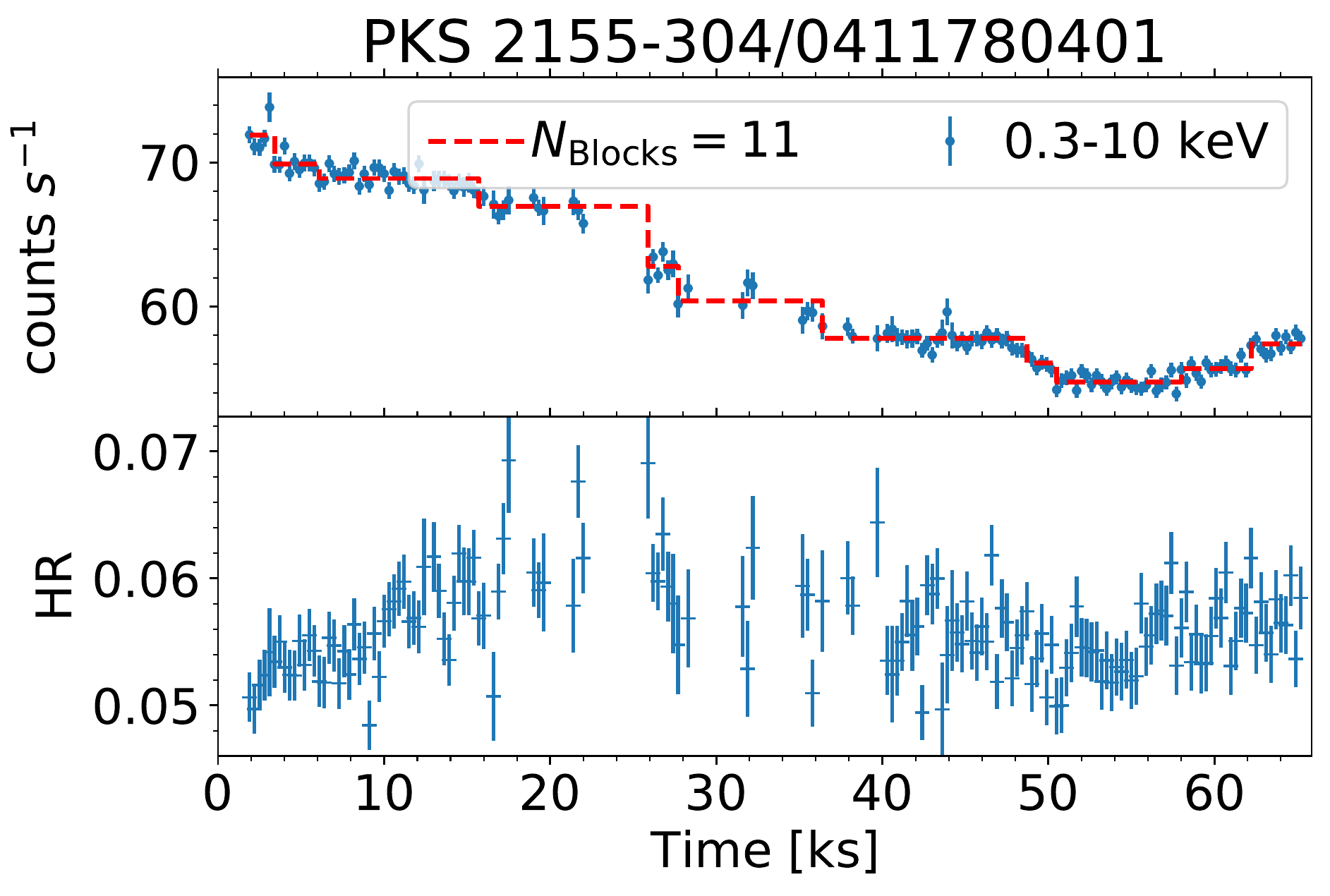}
\includegraphics[scale=0.29]{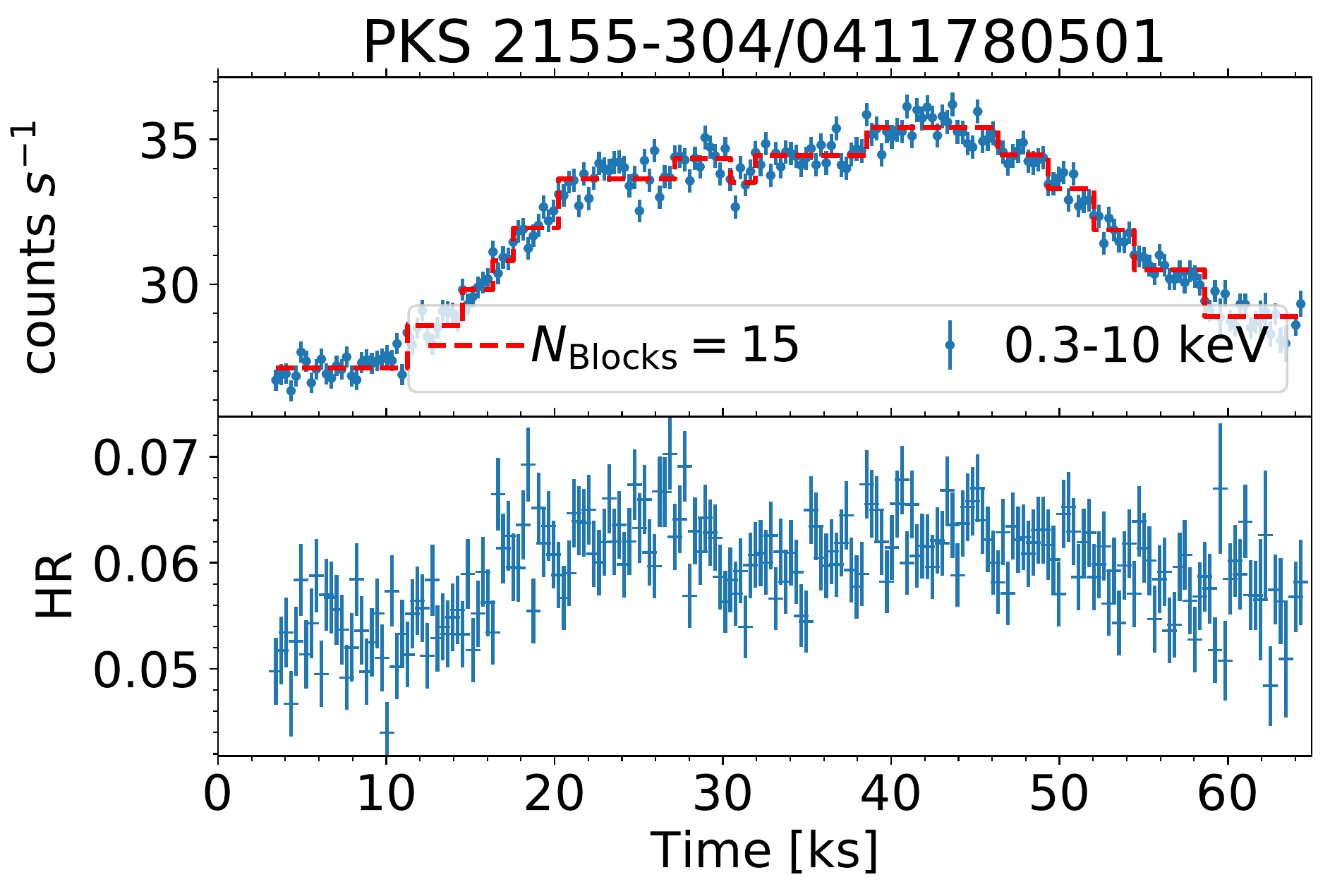}
\includegraphics[scale=0.29]{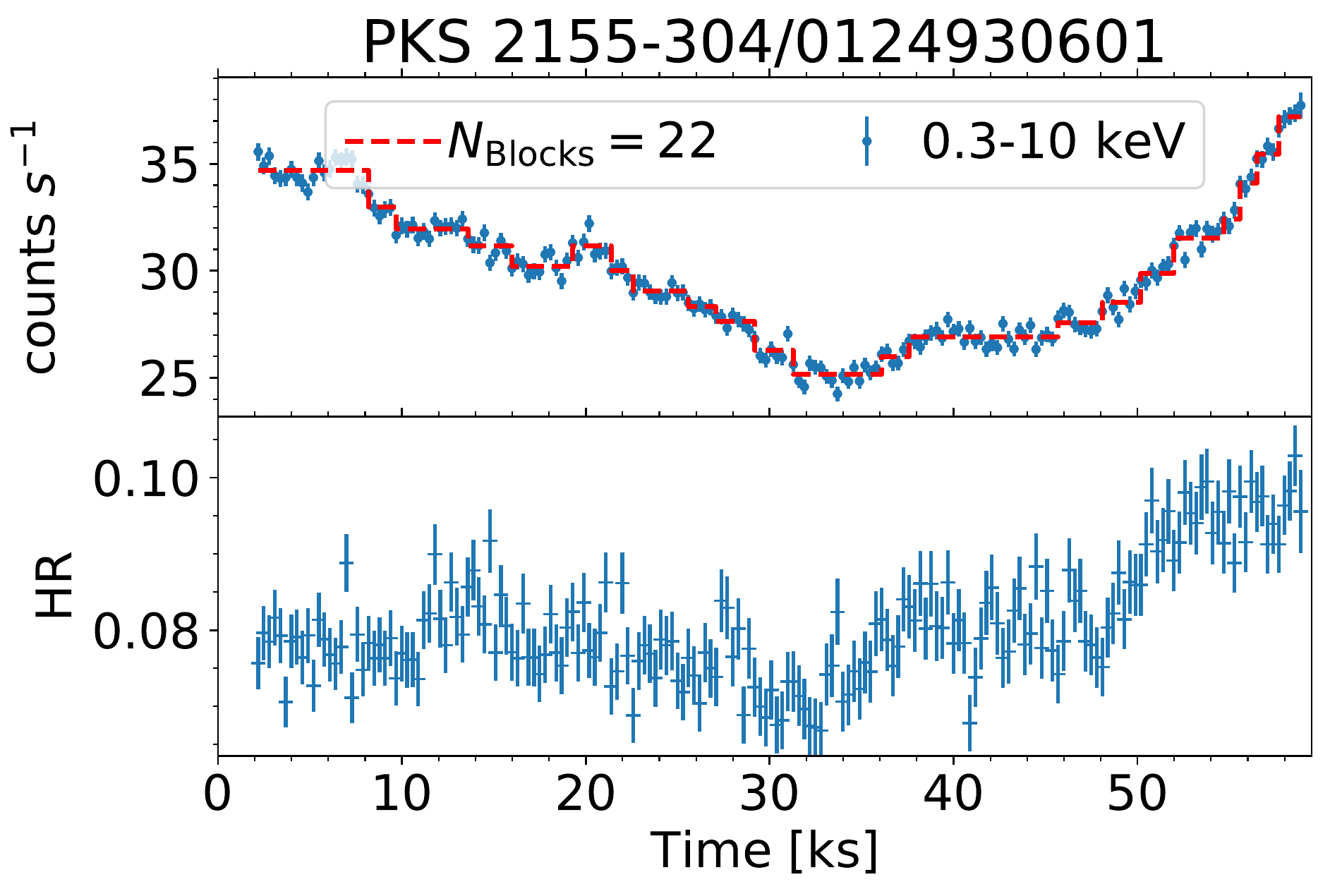}
\includegraphics[scale=0.29]{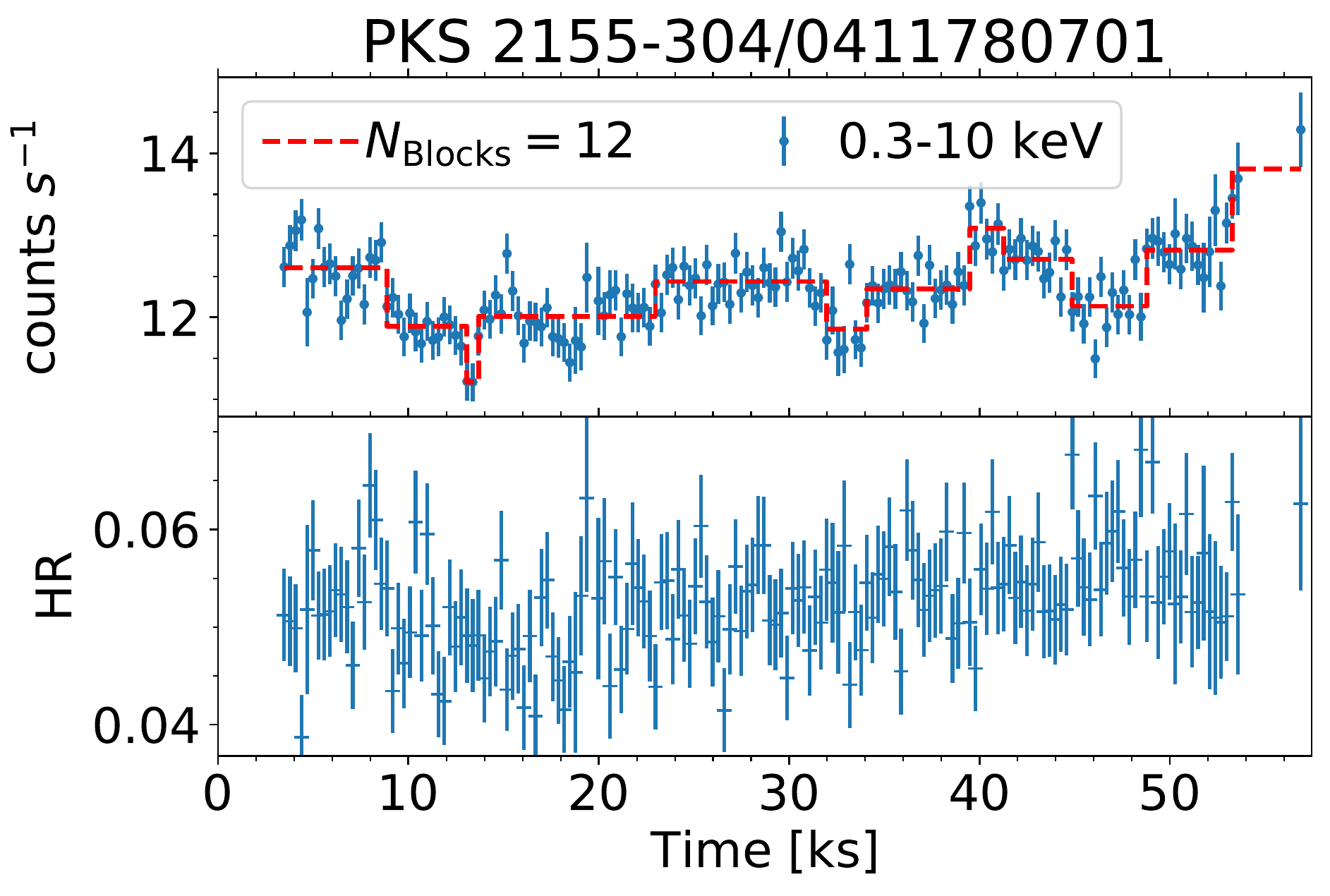}
\includegraphics[scale=0.29]{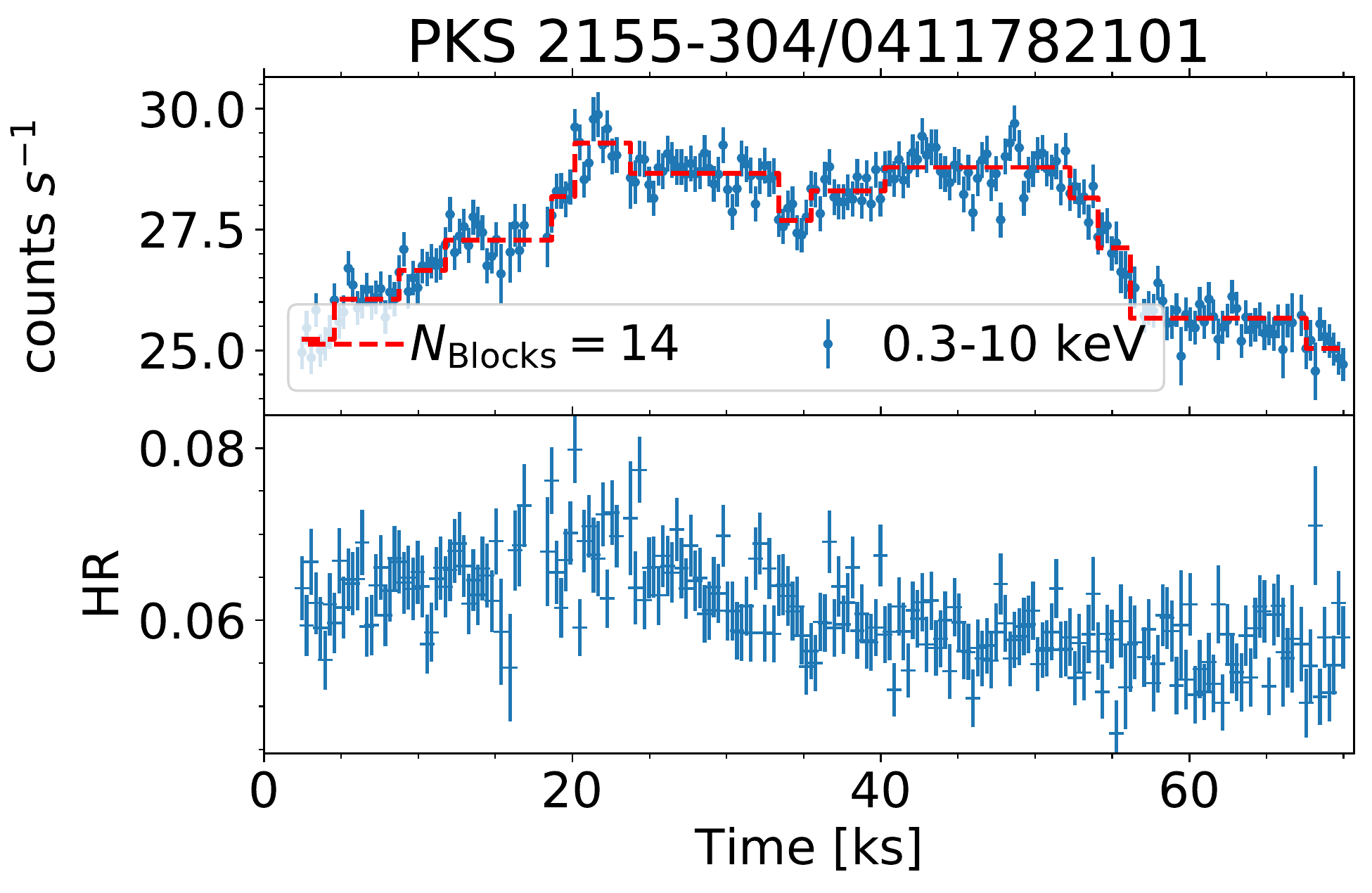}
\includegraphics[scale=0.29]{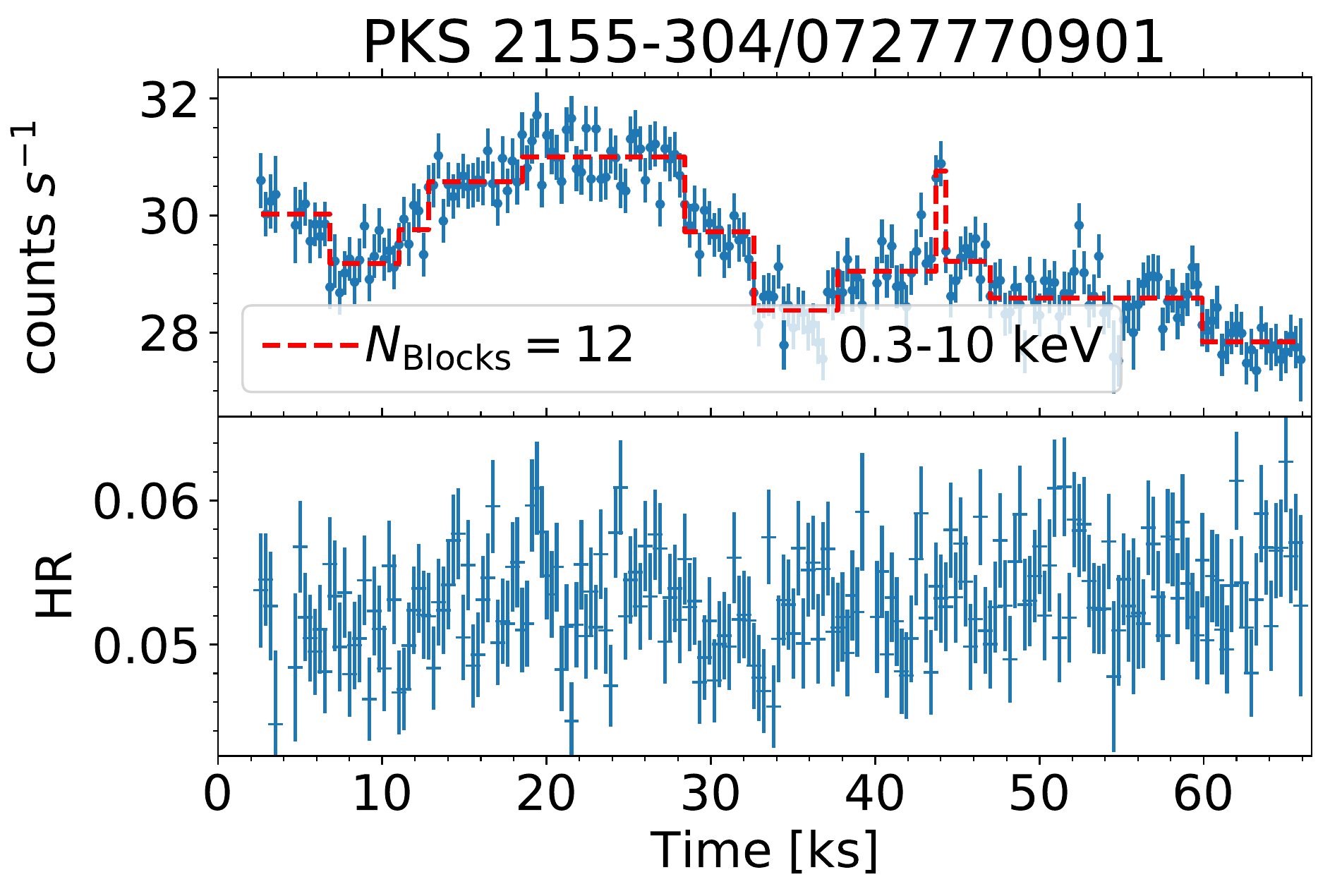}
\includegraphics[scale=0.29]{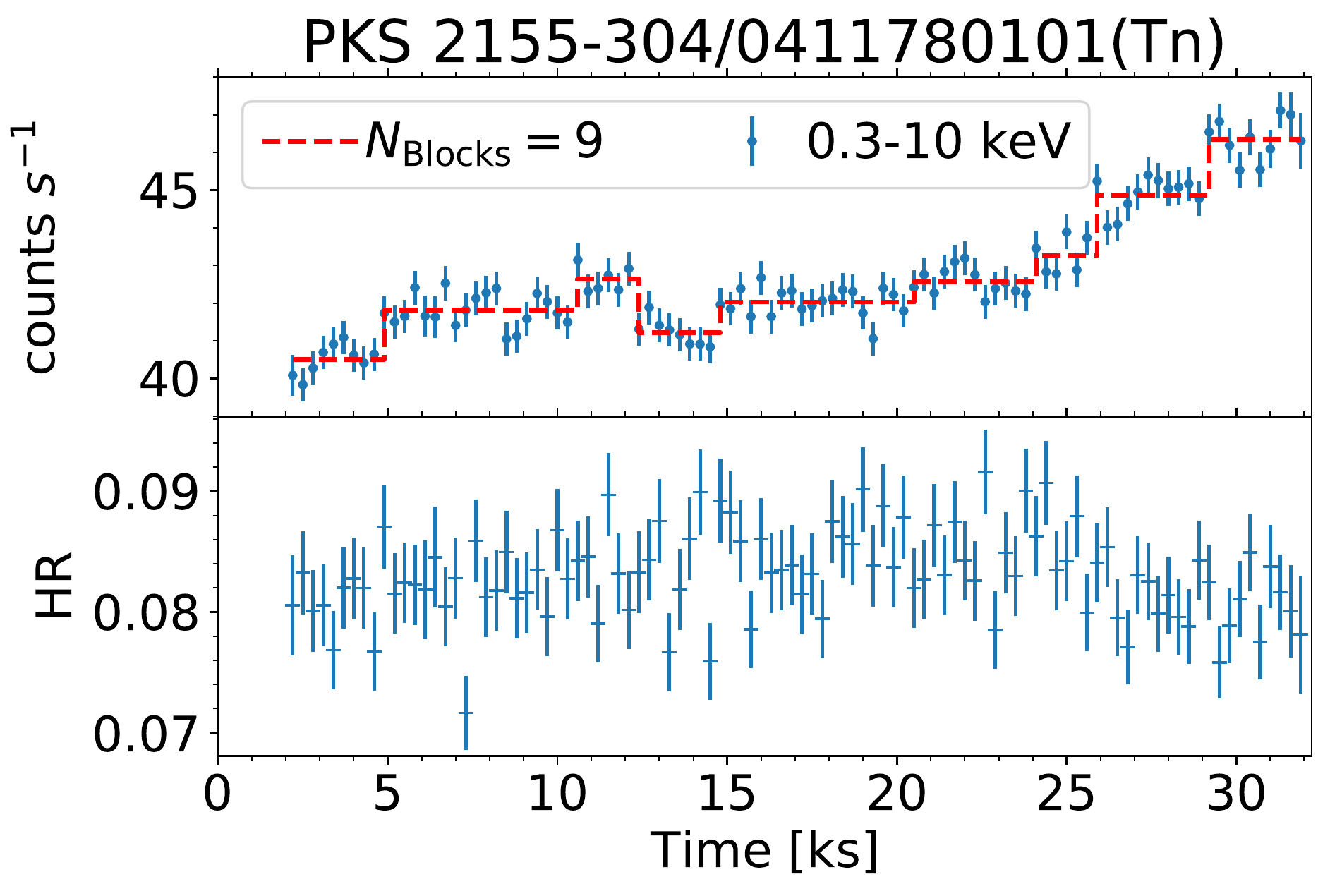}
\includegraphics[scale=0.29]{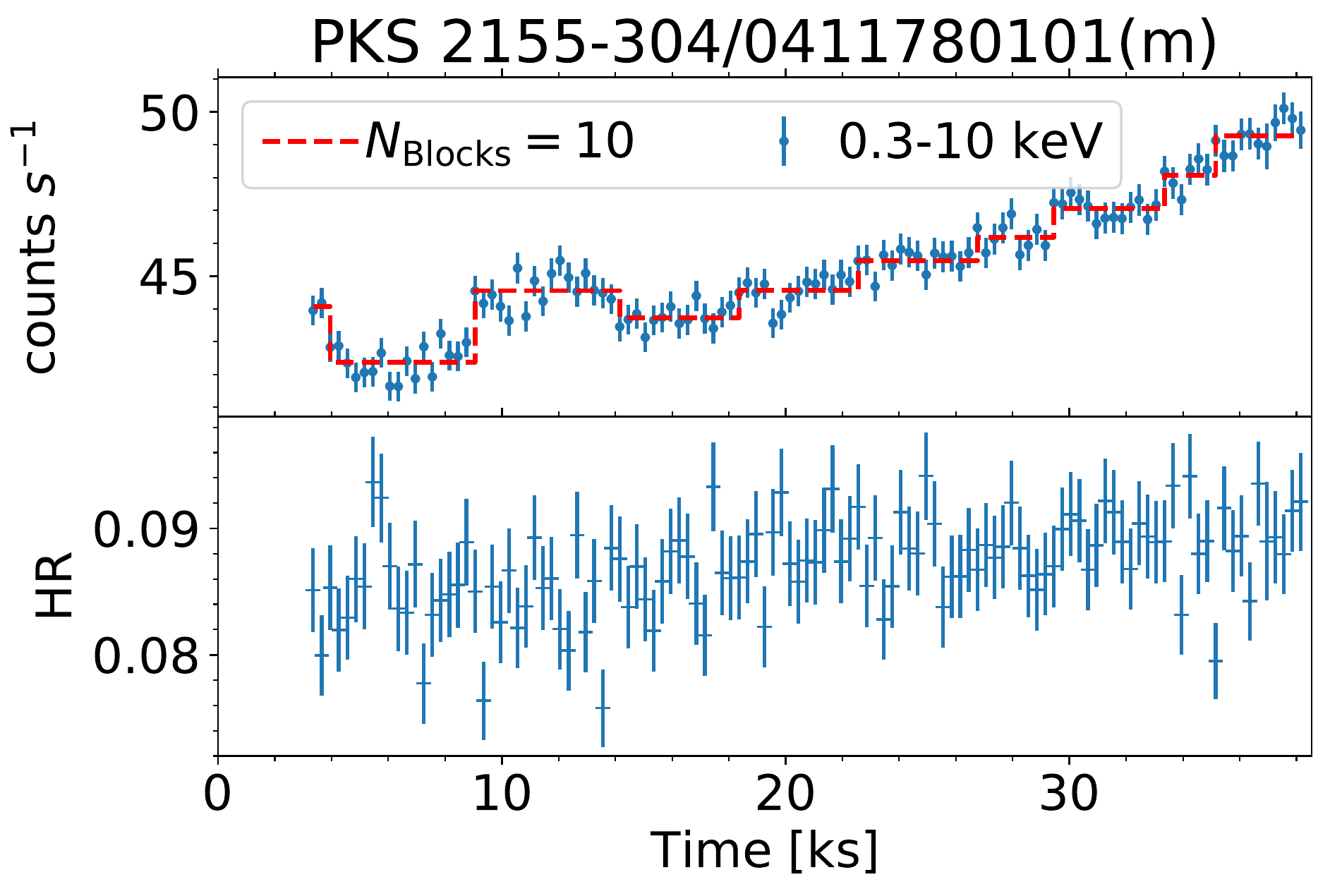}
\includegraphics[scale=0.29]{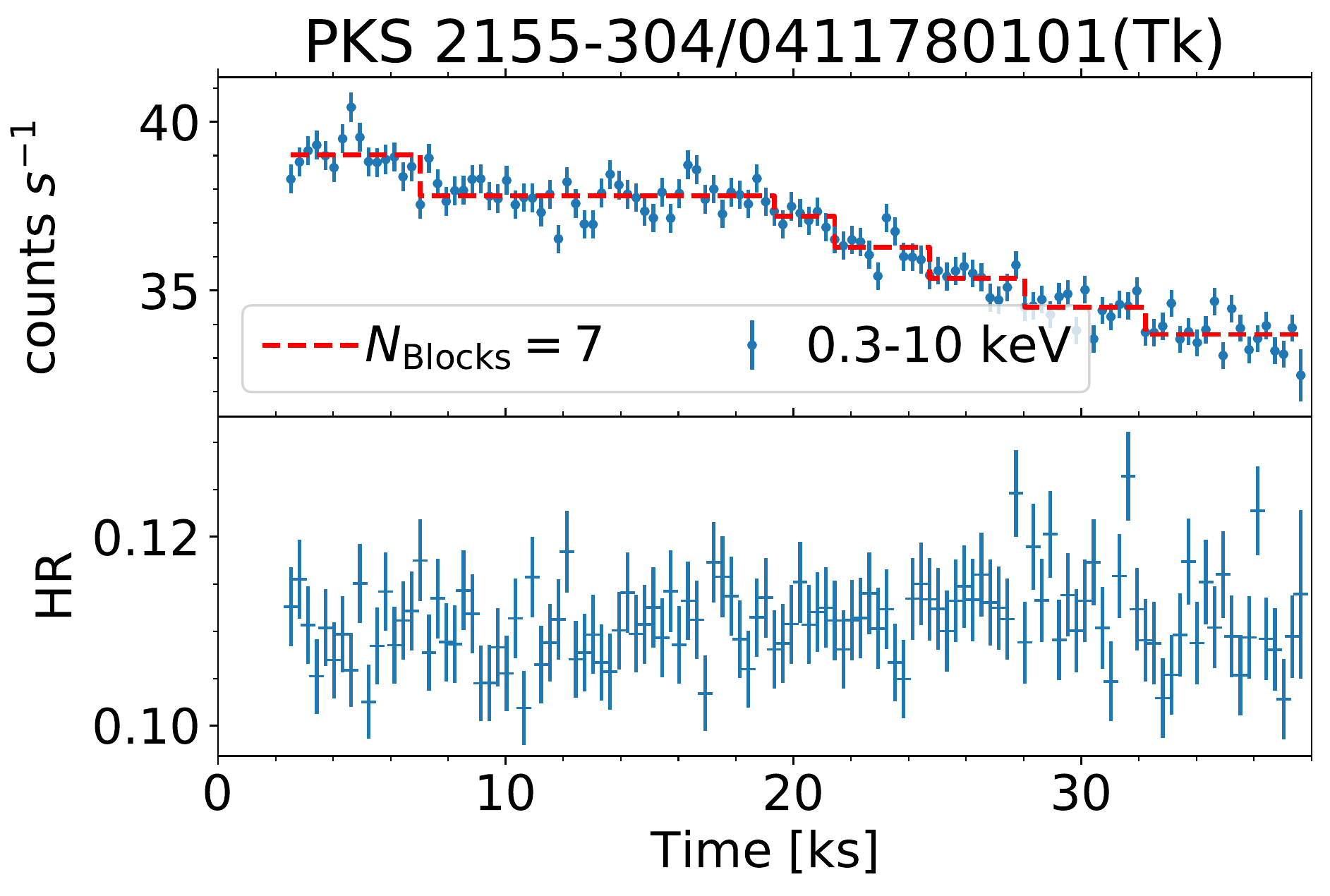}
\caption{{\it XMM-Newton} X-ray light curves of blazars S5 0716+714 (IBL) and PKS
2155-304 (HBL) using 300s binning along with the hardness ratio. The dashed line
shows the Bayesian blocks, marking changes in count rate with time. ``Tn'', ``m'',
and ``Tk'' within the parenthesis in the last three light curves refer to the
filters ``thin'', ``medium'', and ``thick'' respectively (ref Table \ref{tab:obslog}).}
\label{fig:lc}
\end{figure}

\section{Analysis and Results}\label{sec:Analysis}

\subsection{Light curves}
\begin{figure}
\centering
\includegraphics[scale=0.29]{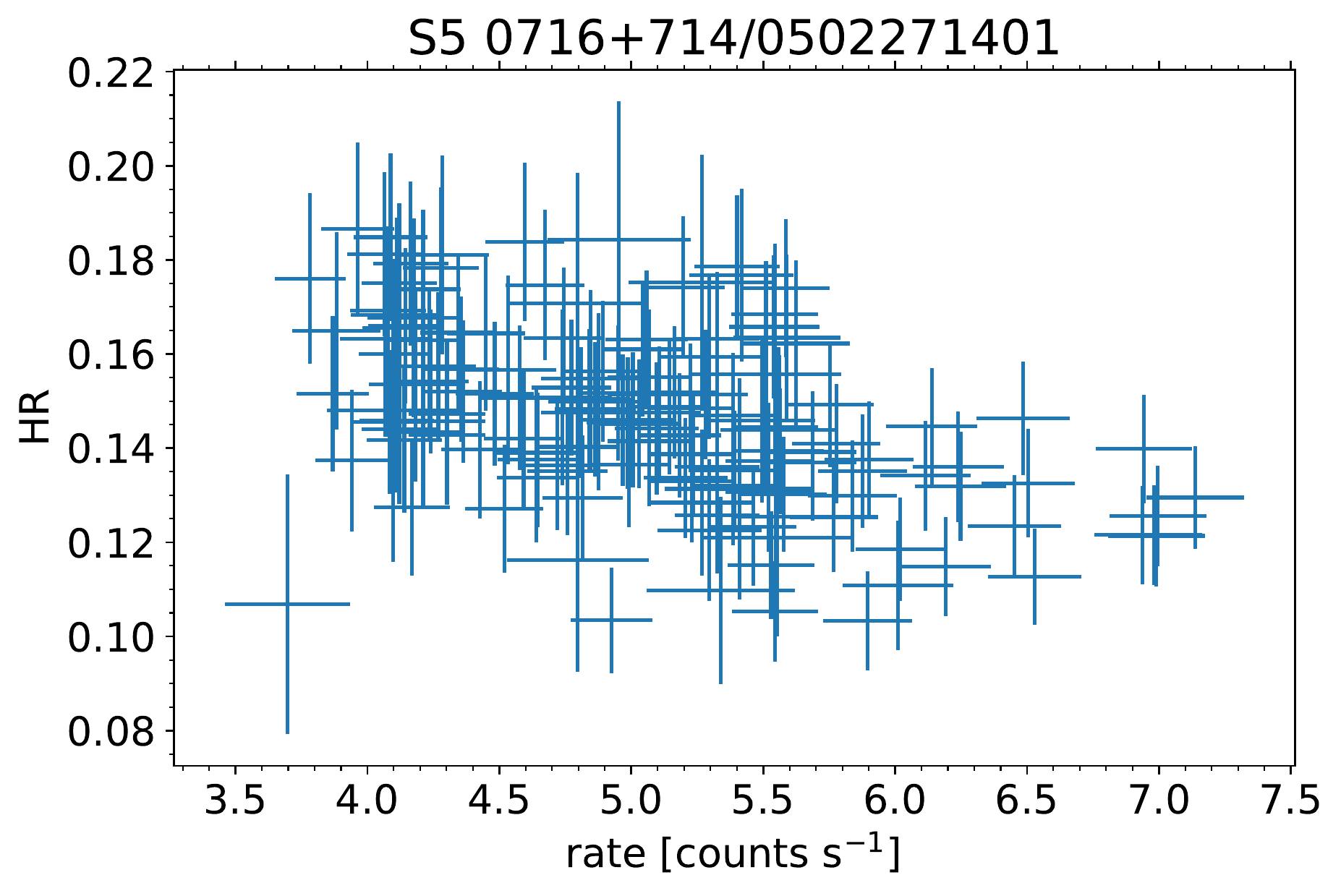}
\includegraphics[scale=0.29]{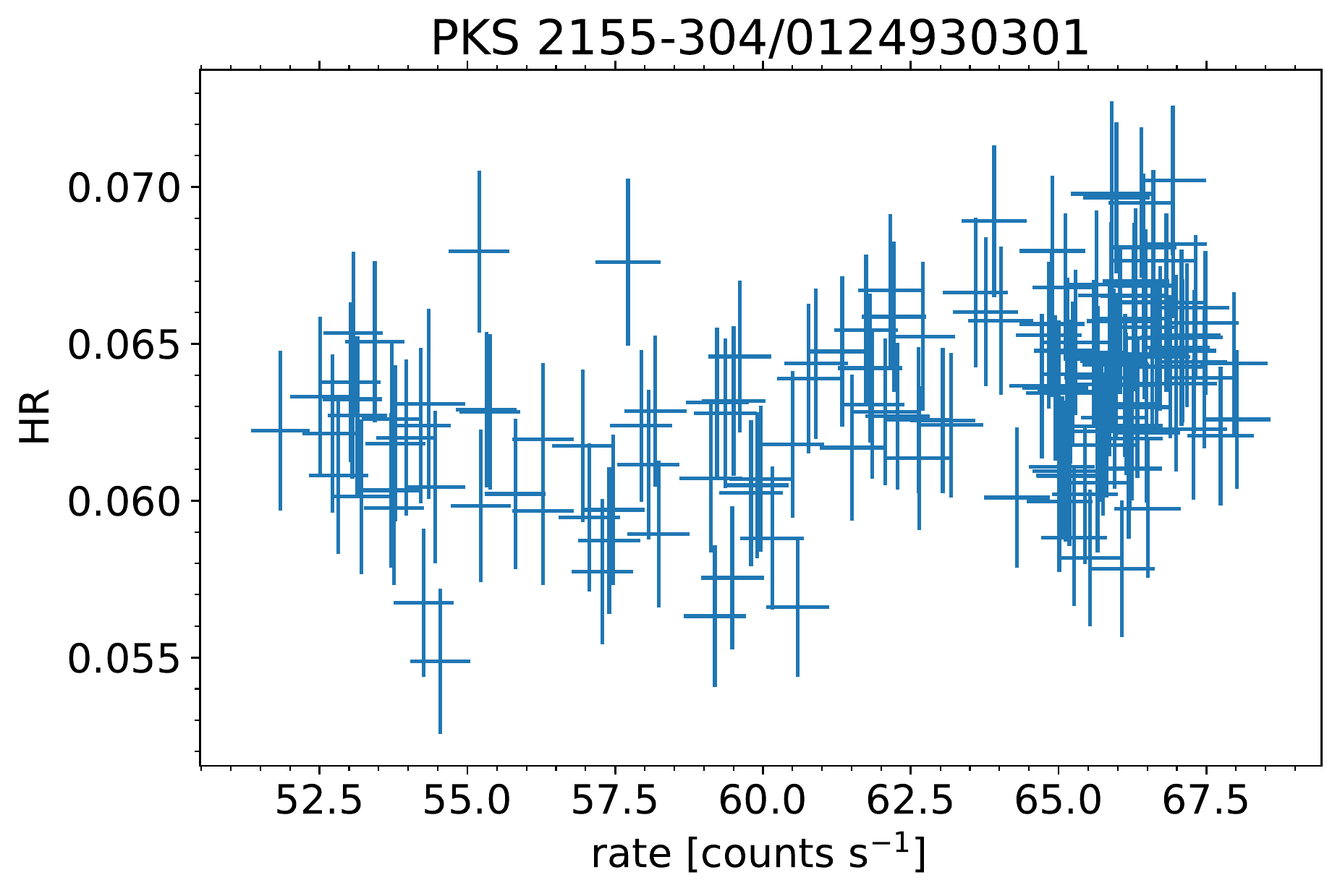}
\includegraphics[scale=0.29]{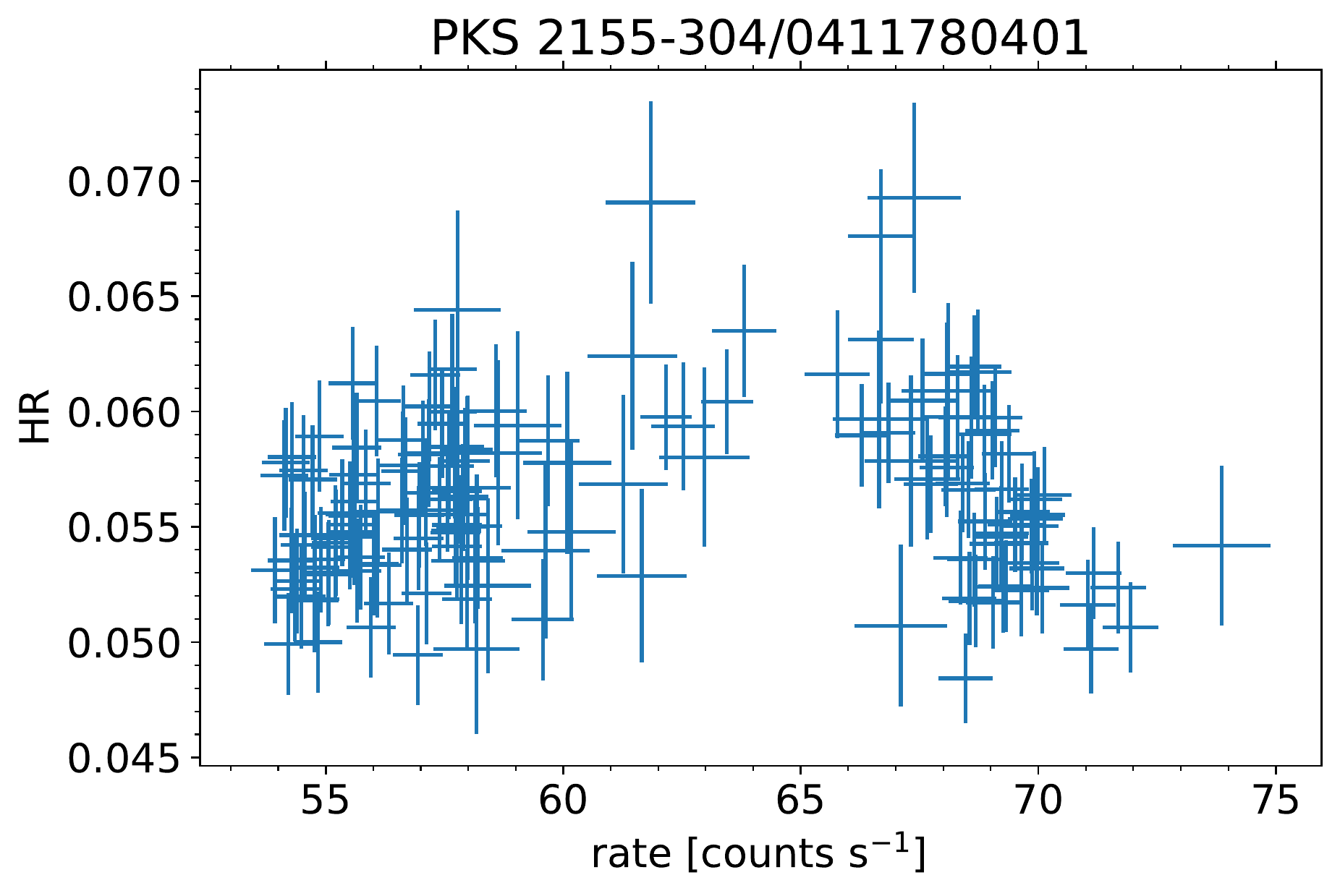}
\includegraphics[scale=0.29]{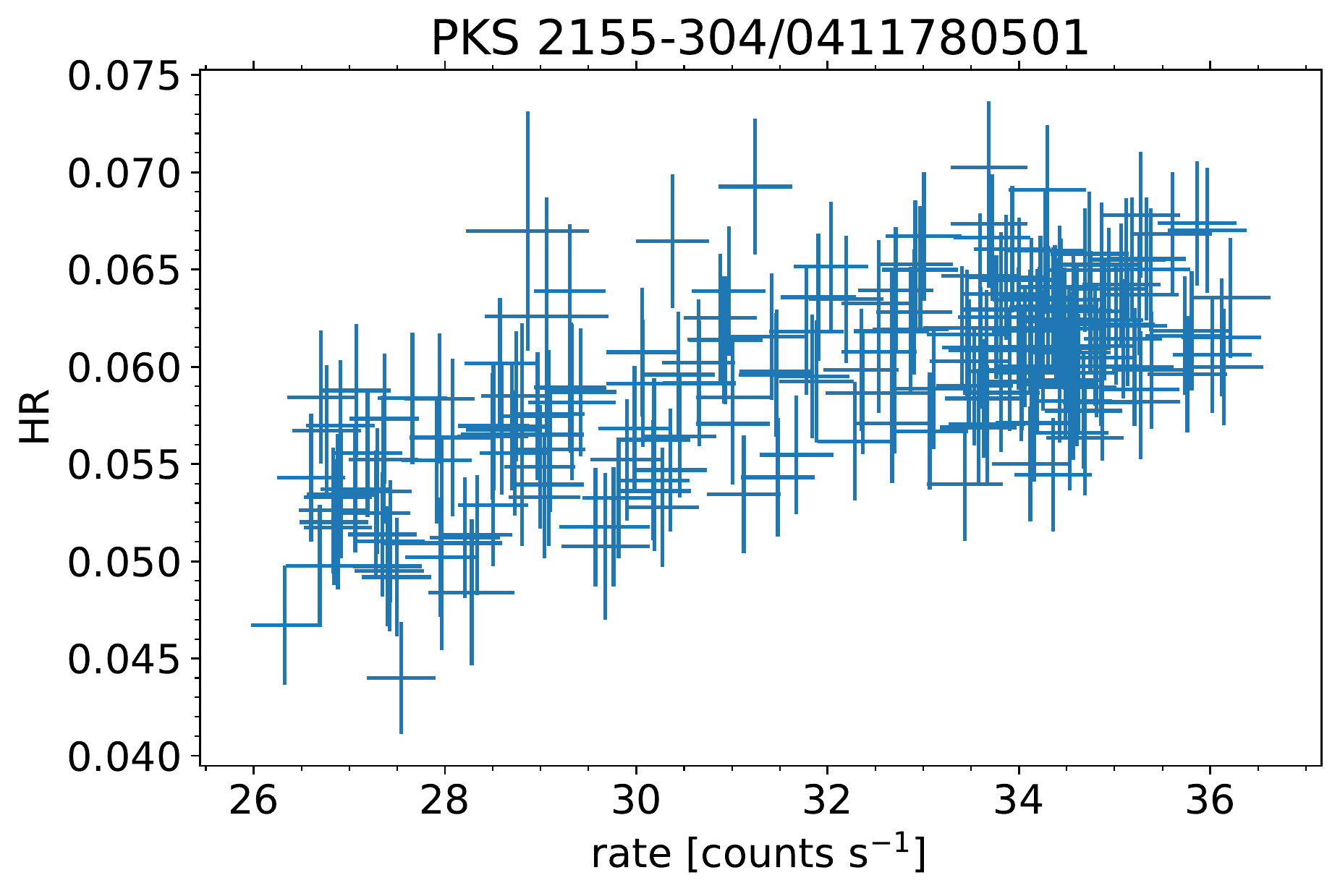}
\includegraphics[scale=0.29]{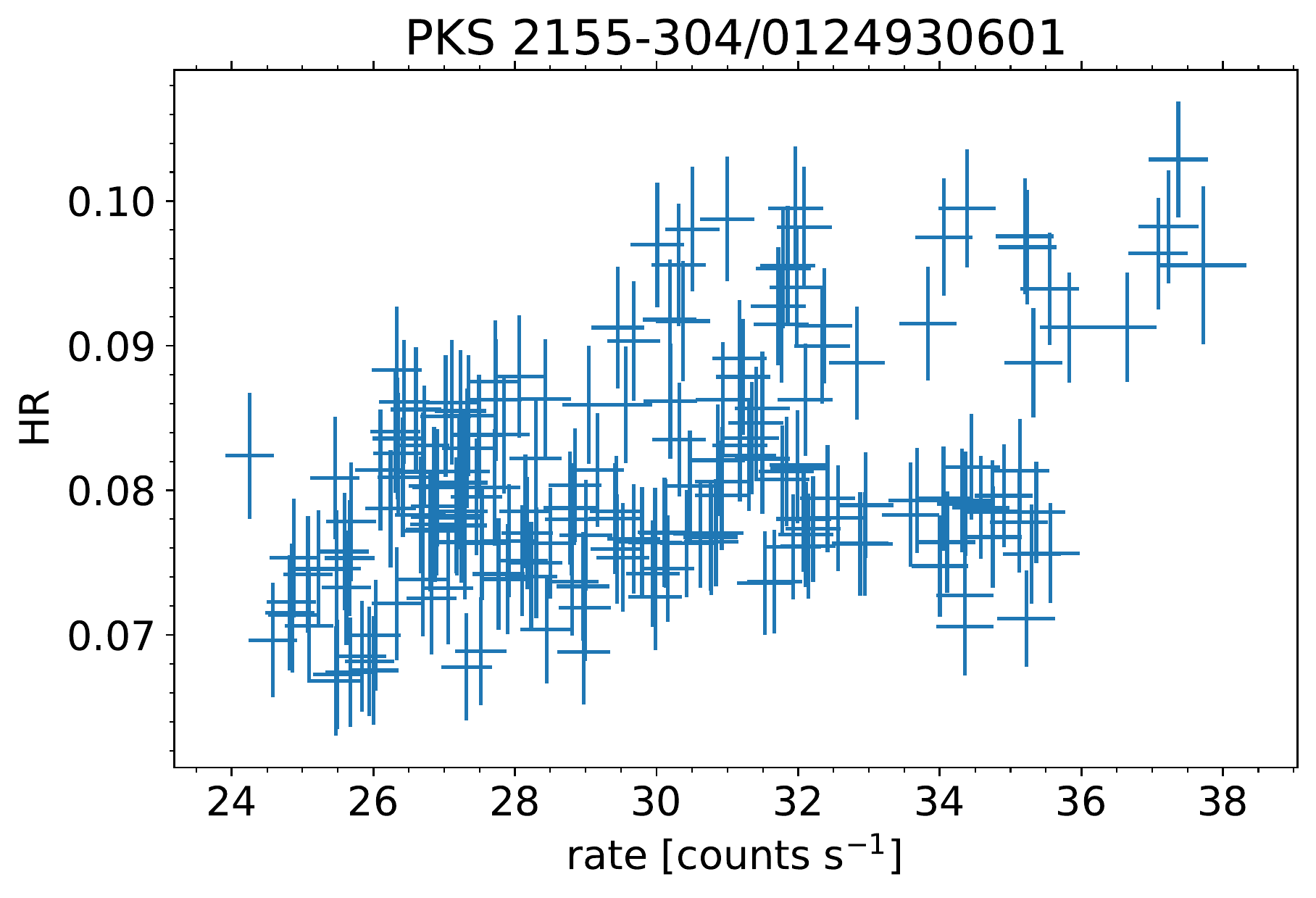}
\includegraphics[scale=0.29]{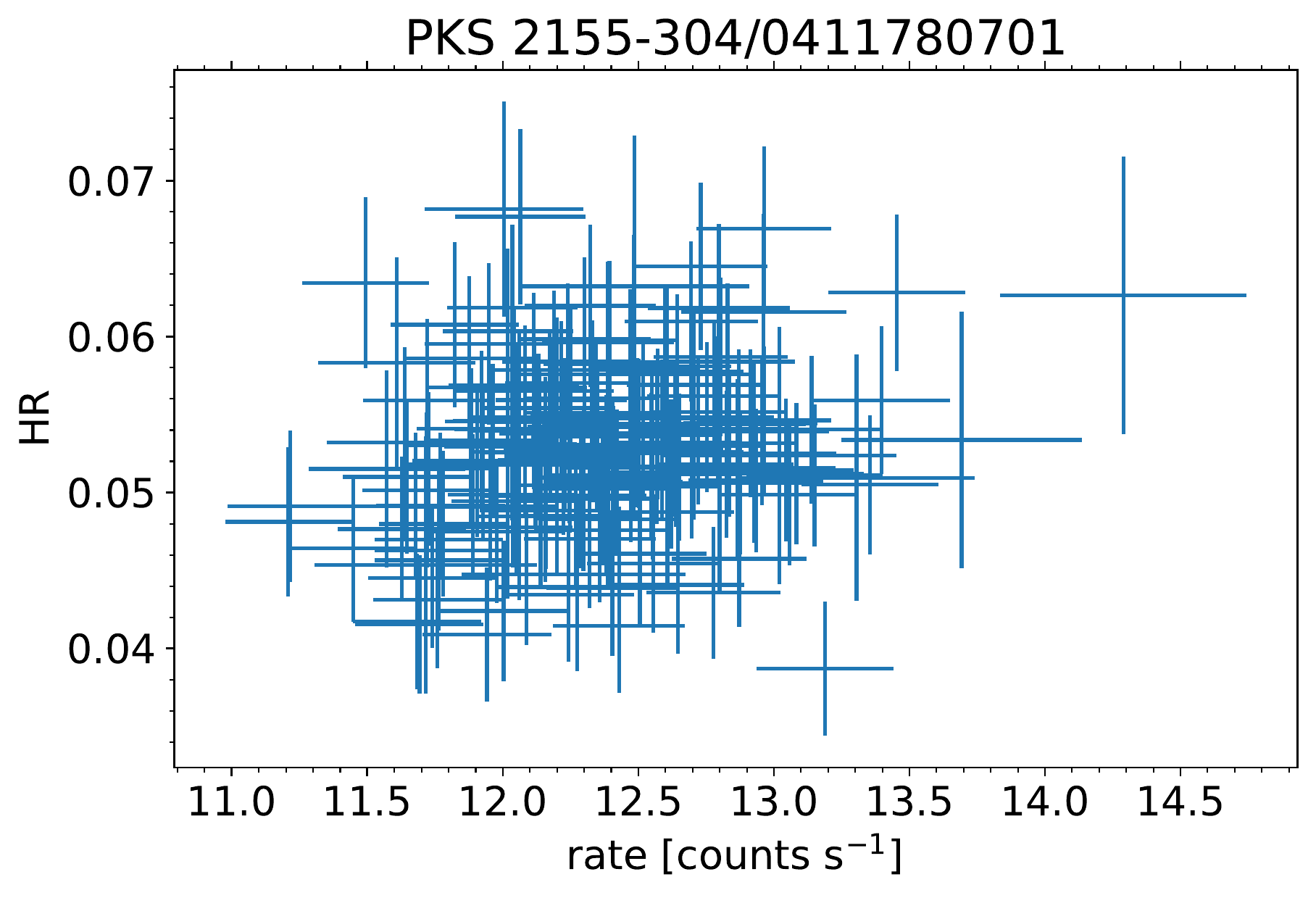}
\includegraphics[scale=0.29]{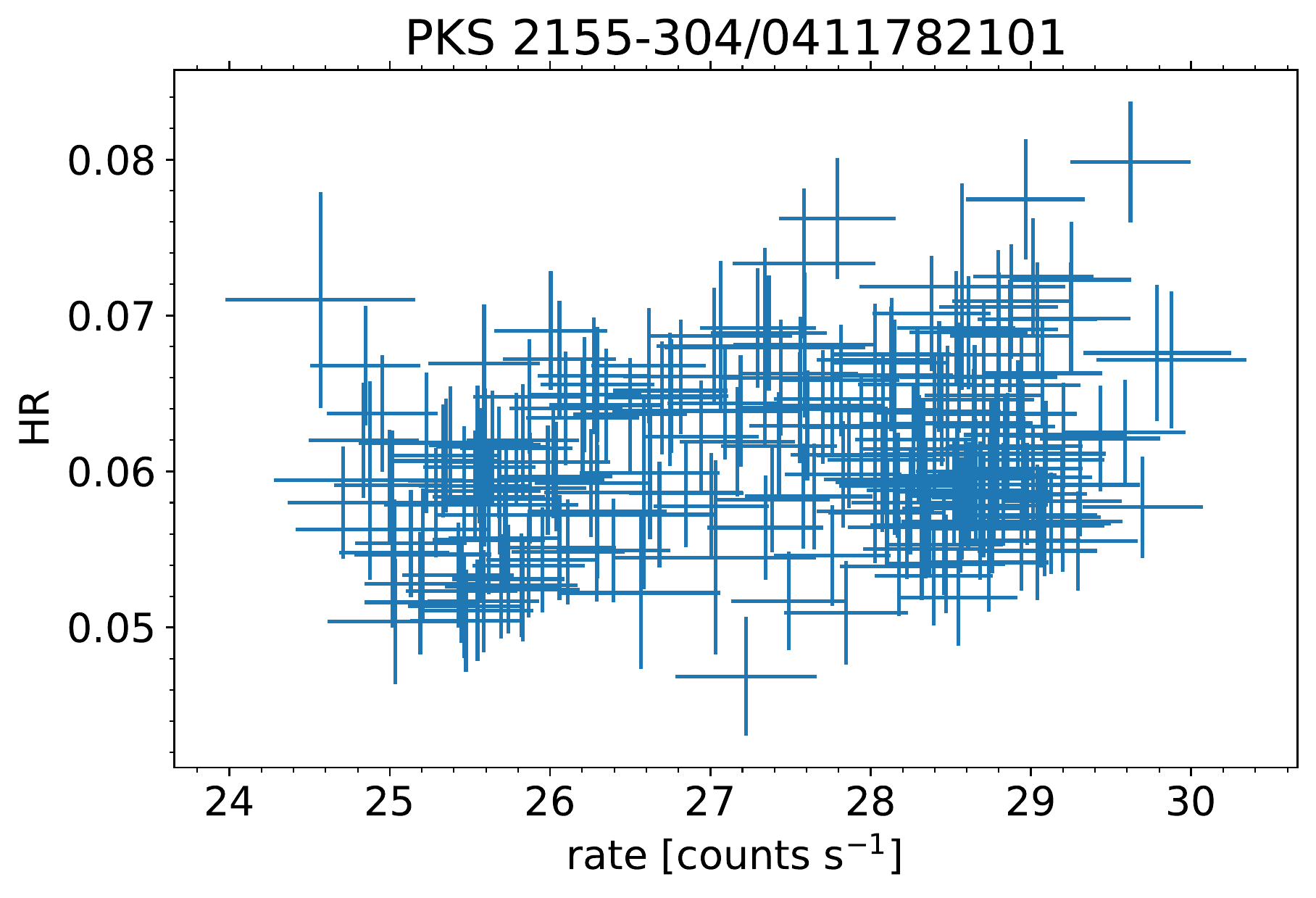}
\includegraphics[scale=0.285]{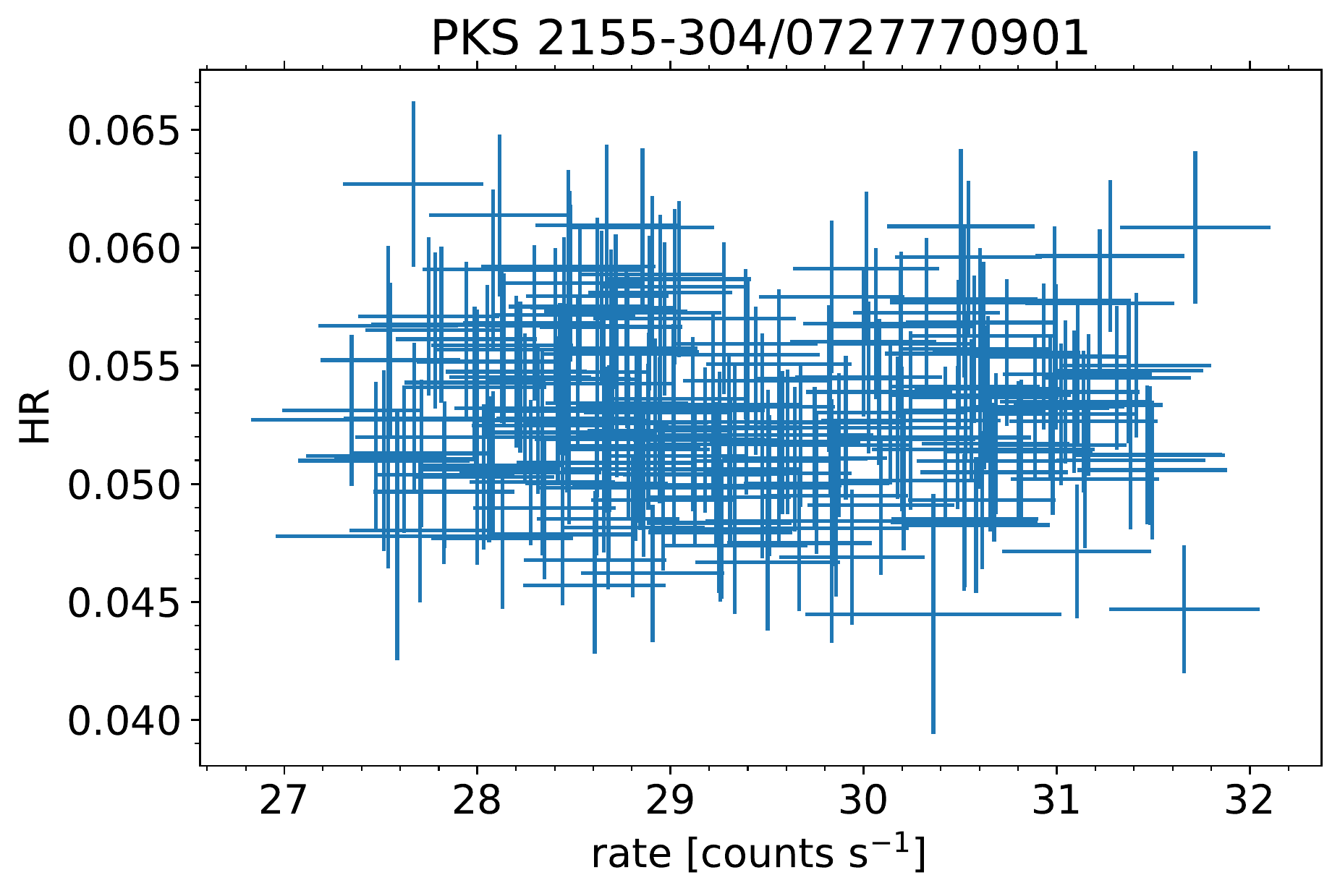}
\includegraphics[scale=0.285]{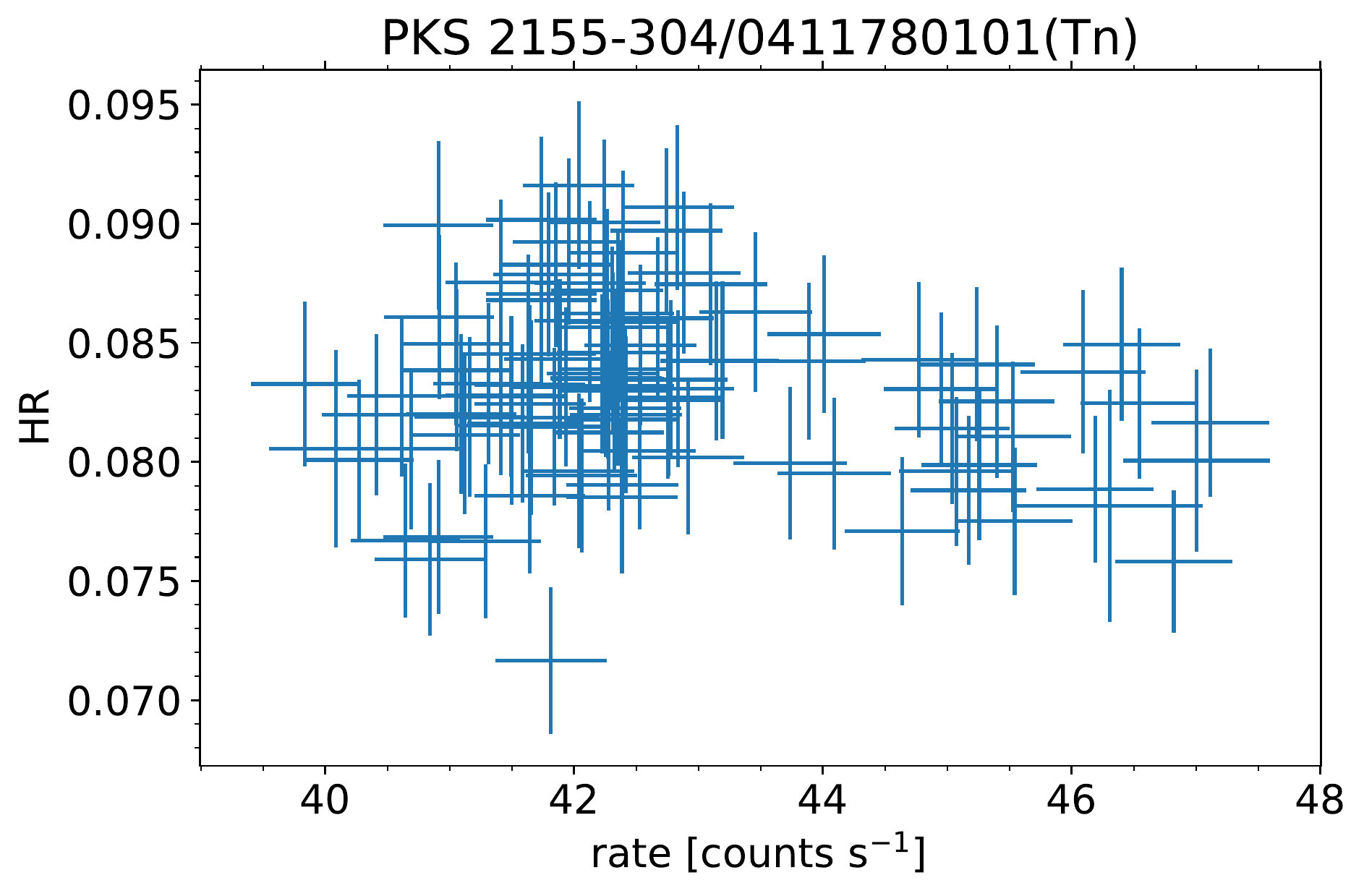}
\includegraphics[scale=0.285]{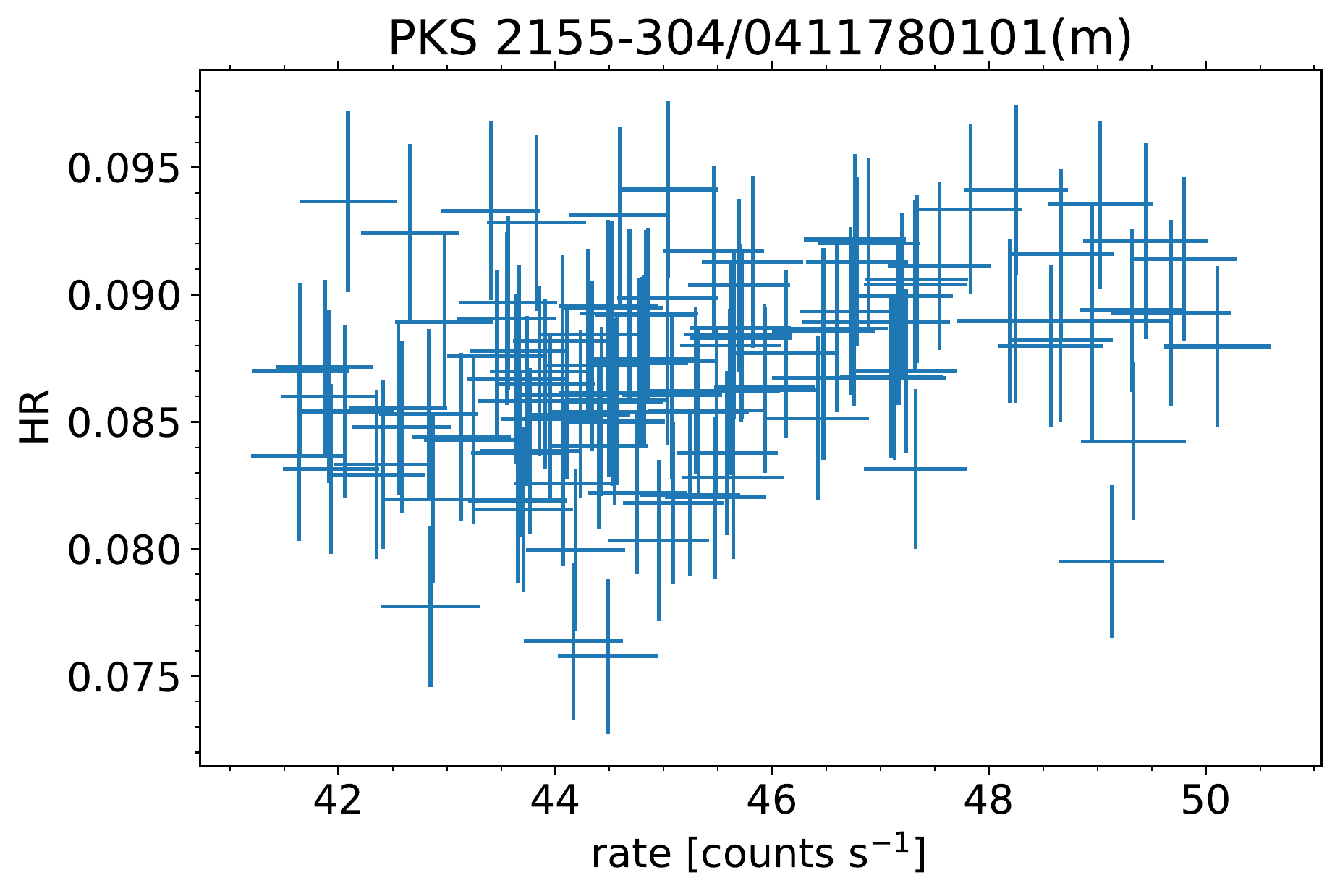}
\includegraphics[scale=0.285]{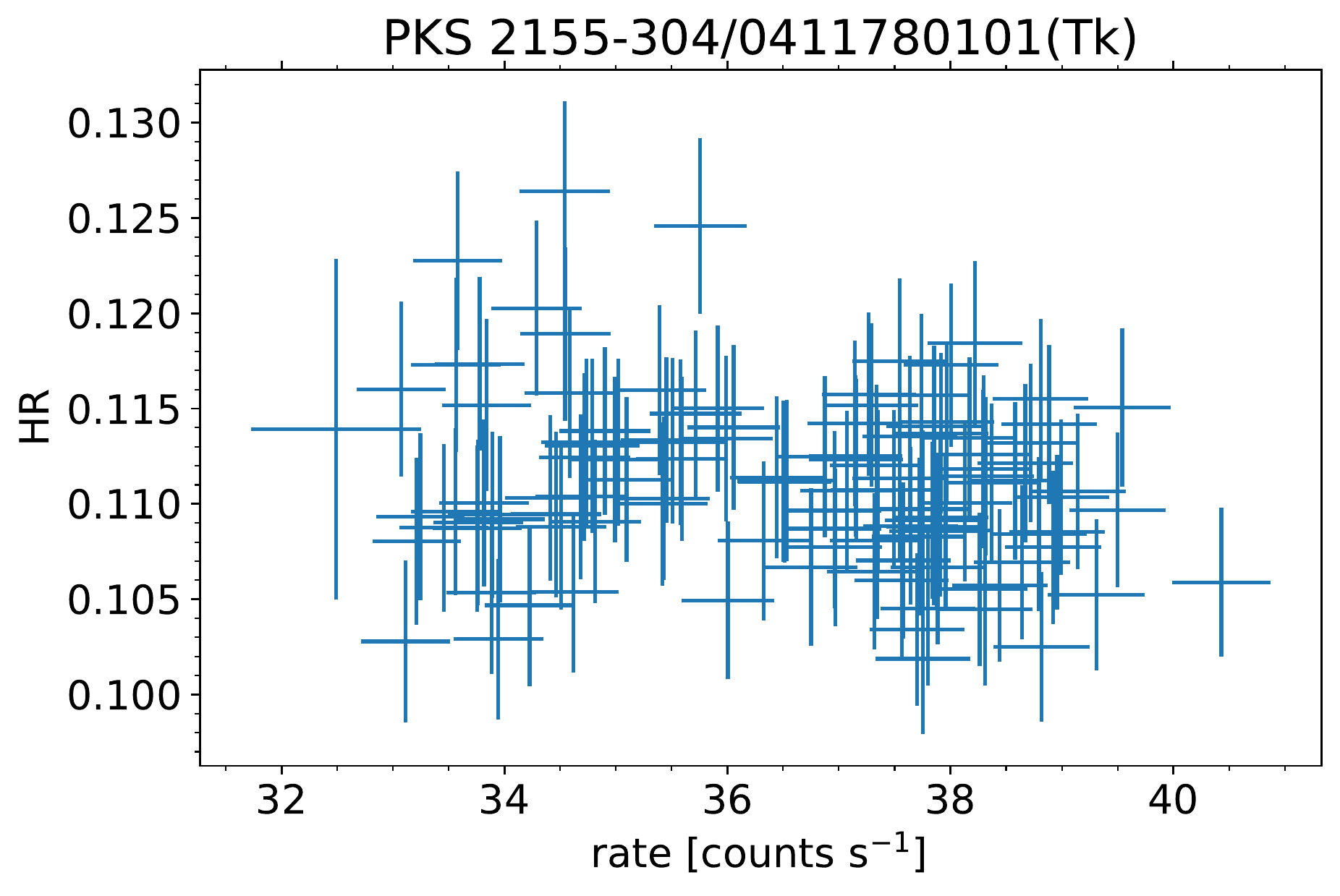}
\caption{HR as a function of 0.3-10 keV count rate.}
\label{fig:HRF}
\end{figure}

The 300s binned light curves along with the corresponding HR for each ID as mentioned
in Table \ref{tab:obslog} are shown in Figure \ref{fig:lc}. The HR here is defined
as the ratio of count rate in the hard energy band i.e. 2 - 10 keV to the count
rate in the soft energy band i.e 0.3 - 2 keV. Additionally, we
have shown the Bayesian blocks \citep{2013ApJ...764..167S}, marking significant changes
in the count rate evolution with time corresponding to p = 0.075. As already stated
above, the 300s binning is employed so that we have enough data for statistical
purpose and yet the trends or correlation are discernible that appears in light
curves binned over a larger duration, e.g. 1000s. Figure \ref{fig:HRF} shows HR as
a function of the count rate in the 0.3 -- 10 keV band. 

Except for the only ID corresponding to the IBL S5 0716+714, the rest of the light
curves shown in Figure \ref{fig:lc} belong to PKS 2155-304 and capture X-ray state
from very low -- characterized by low count rate to a high state with a rate more
than five times higher. The change in count rate is also different during
different states and ranges between 1.2 -- 1.5 (ref also Figs \ref{fig:hist} and
\ref{fig:rmsF}). Further, except for the three IDs: 
0411780301, 0411780401, and 0411780101(Tn) of PKS 2155-304, the rest of the IDs
shows correlated variability of the count rate with HR -- increase in count rate
accompanied by an increase of HR and vice-versa. These three,
on the other hand, show complex behavior with respect to HR. For some periods,
the count rate and HR are correlated while anti-correlated -- high count rate but low
HR and vice-versa, for other periods. This feature is also visible in the HR plot for these
IDs shown in Figure \ref{fig:HRF}. It should, however, be noted that the sample size
is biased, as already mentioned previously that we are interested in observations
exhibiting variability and have good photon statistics. Interestingly, of these three
IDs showing complex trends, the first two (0411780301, 0411780401) are the brightest
of the current sample.

For S5 0716+714, we have only one observation corresponding to a high state of the
source and shows count rate changes almost by a factor of two and it anti-correlates
with HR.

In Table \ref{tab:test}, we show the result of the test of stationarity using the Augmented Dickey-Fuller (ADF) test and the Kwiatkowski-Phillips-Schmidt-Shin (KPSS)
test. The parenthesis marks the result for the light curve in logarithm units. Both
method checks for ``unit root'' in the time series. The former checks for a
``unit root'' and difference stationarity in the time series with the null hypothesis
that ``a unit root exist'' i.e. series is non-stationary. The latter i.e. the KPSS 
test, on the other hand, checks for stationarity around a deterministic trend but
its null hypothesis is the opposite of the ADF i.e. the series is stationary. For both
the tests, \(p \leq 0.05\) was chosen to reject the null hypothesis. Thus, in the
case of ADF, rejection of the null hypothesis means that the time series is stationary
while for KPSS it means a non-stationary time series.

\begin{table}[H]
\caption{Results of Stationarity test using ADF and KPSS}
\centering
\begin{tabular}{cccccc}
\toprule
\textbf{Source}	& \textbf{Observation-ID} & rate (count s$^{-1}$) & ADF & KPSS & AD$^*$\\
\midrule
S5~0716+714	& 0502271401  &5.0$\pm 0.6$  & N (N) & S (S) & log-normal\\
PKS~2155--304& 0124930301 &62.2$\pm 0.1$ & N (N) & N (N) & -\\
            & 0411780401 &61.2$\pm 0.1$  & N (N) & N (N) & -\\
            & 0411780501 &31.60$\pm 0.03$& N (N) & N (N) & -\\
            & 0124930601 &29.8$\pm 0.03$ & N (N) & N (N) & -\\
            & 0411780701 &11.82$\pm 0.02$& N (N) & N (N) & Normal/log-normal\\ 
            & 0411782101 &24.40$\pm 0.11$& N (N) & N (N) & -\\
            & 0727770901 &29.40$\pm 0.03$& N (N) & N (N) & -\\
            & 0411780101 &42.7$\pm 0.1$  & N (N) & N (N) & -\\
            &  &45.3$\pm 0.1$ & N (N) & N (N) & Normal/log-normal\\
            &  &36.5$\pm 0.1$ & N (N) & N (N) & -\\
\bottomrule
S: Stationary; N: Non-stationary\\
*: p=0.01\\
\end{tabular}
\label{tab:test}
\end{table}

\subsection{Histogram}
\begin{figure}[H]
\centering
\includegraphics[scale=0.27]{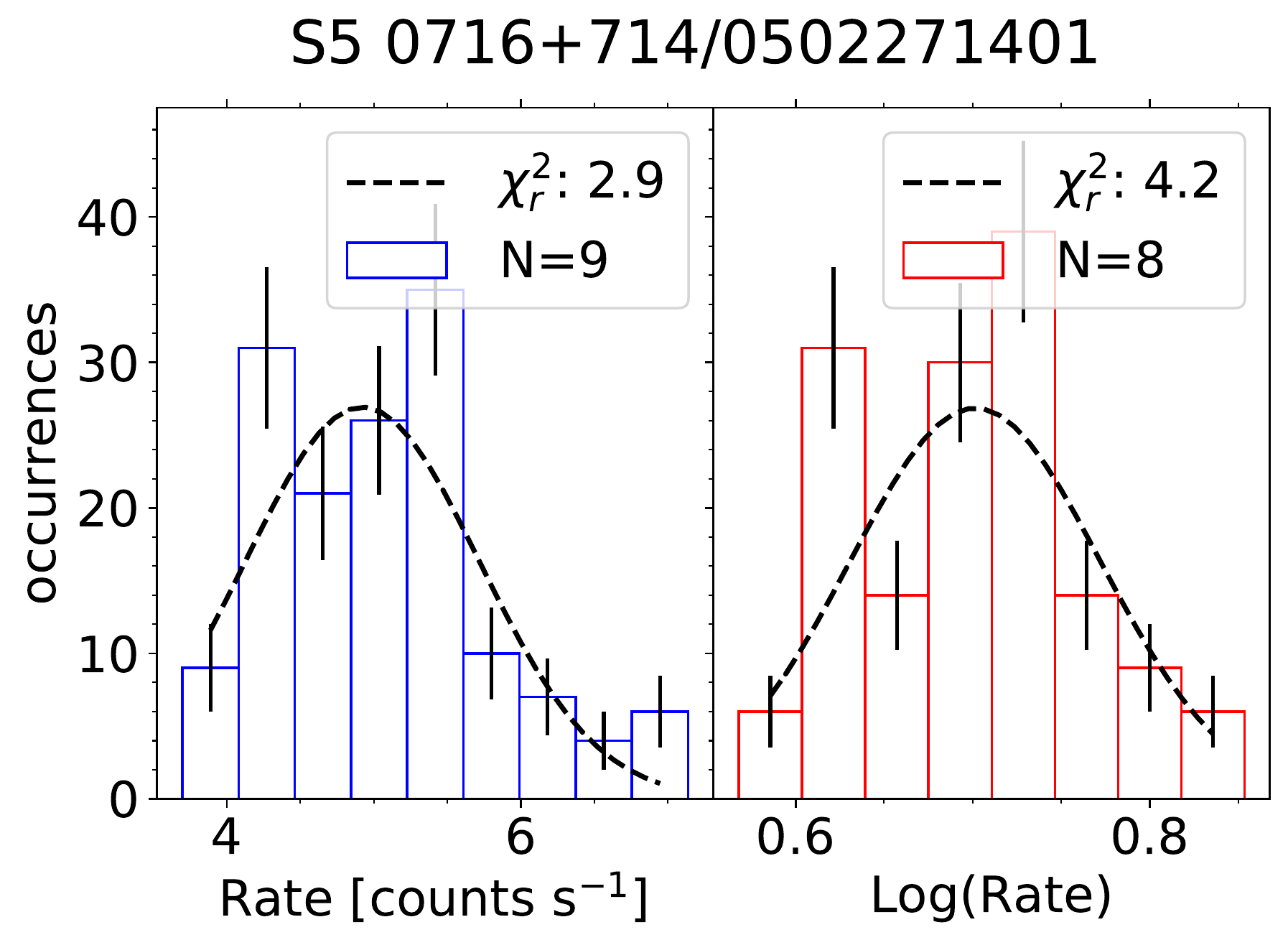}
\includegraphics[scale=0.27]{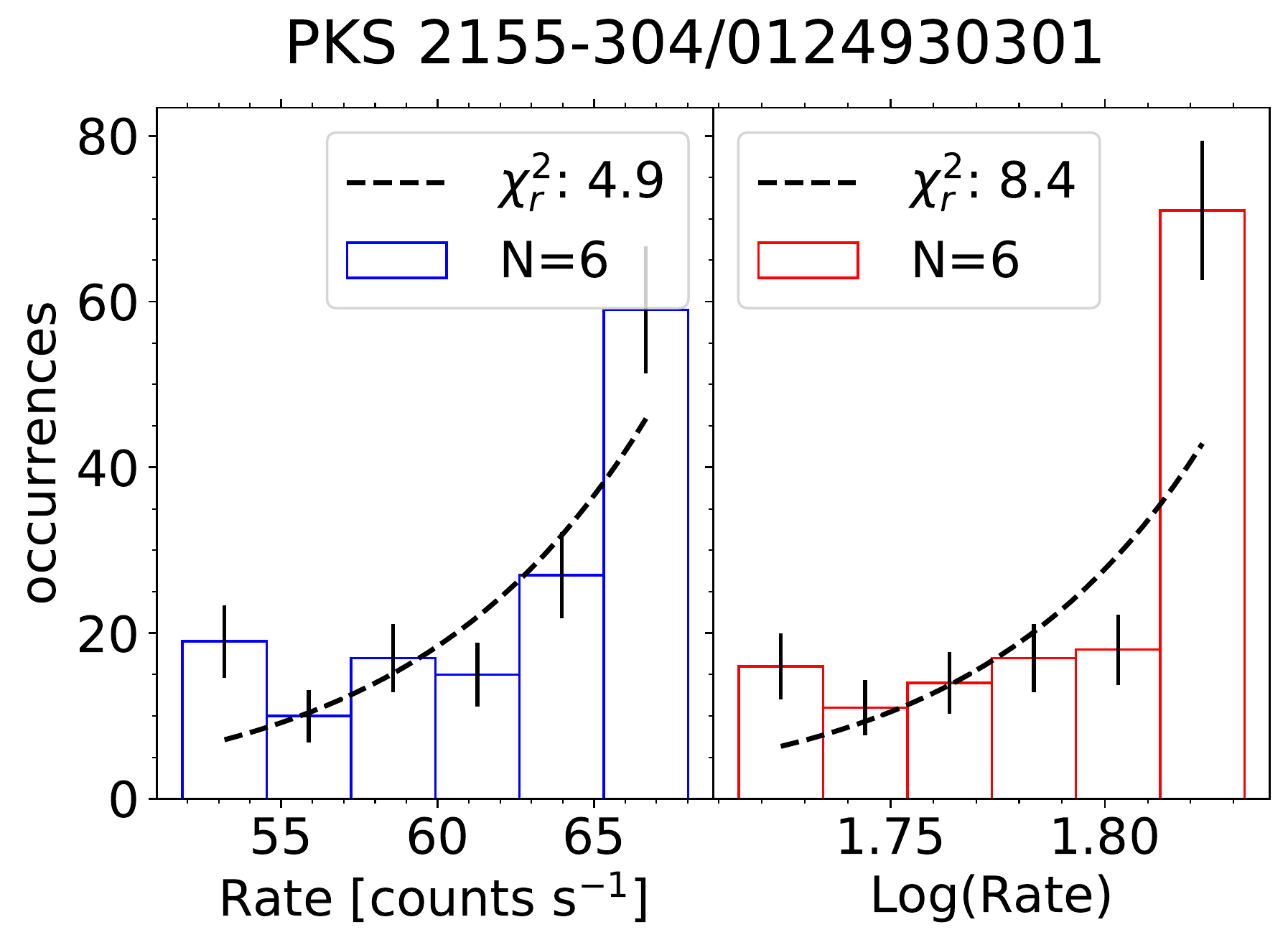}
\includegraphics[scale=0.28]{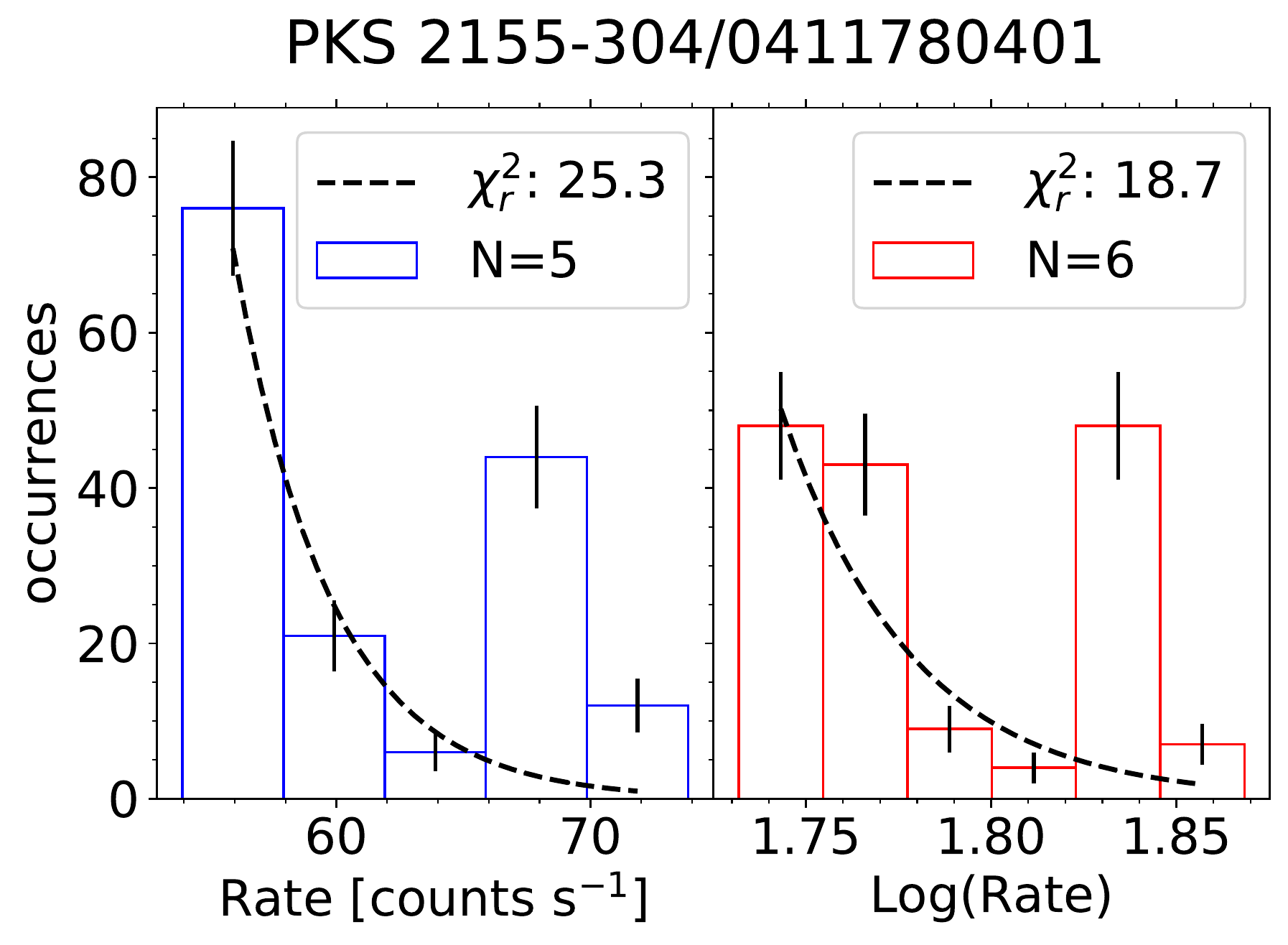}
\includegraphics[scale=0.28]{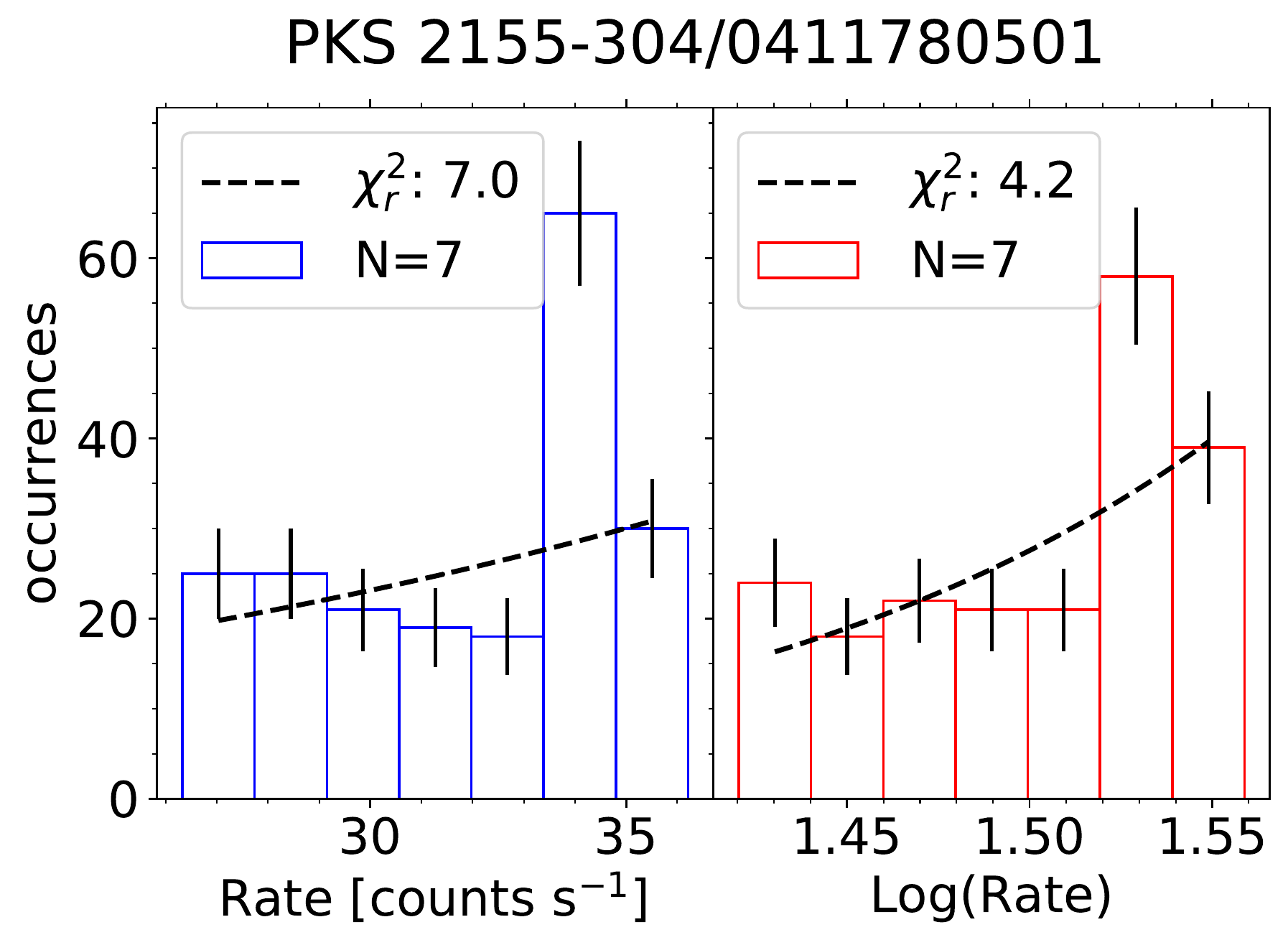}
\includegraphics[scale=0.28]{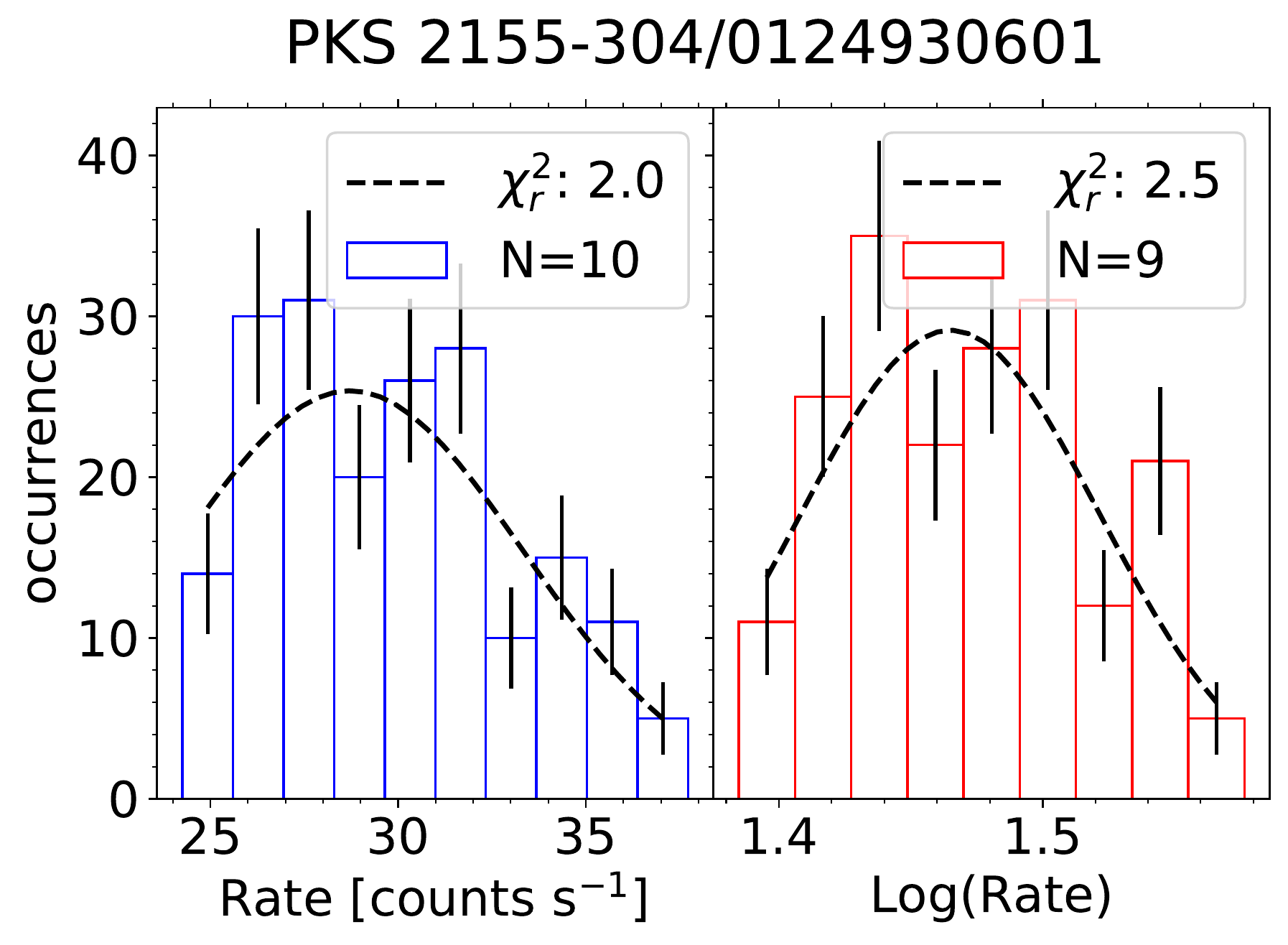}
\includegraphics[scale=0.275]{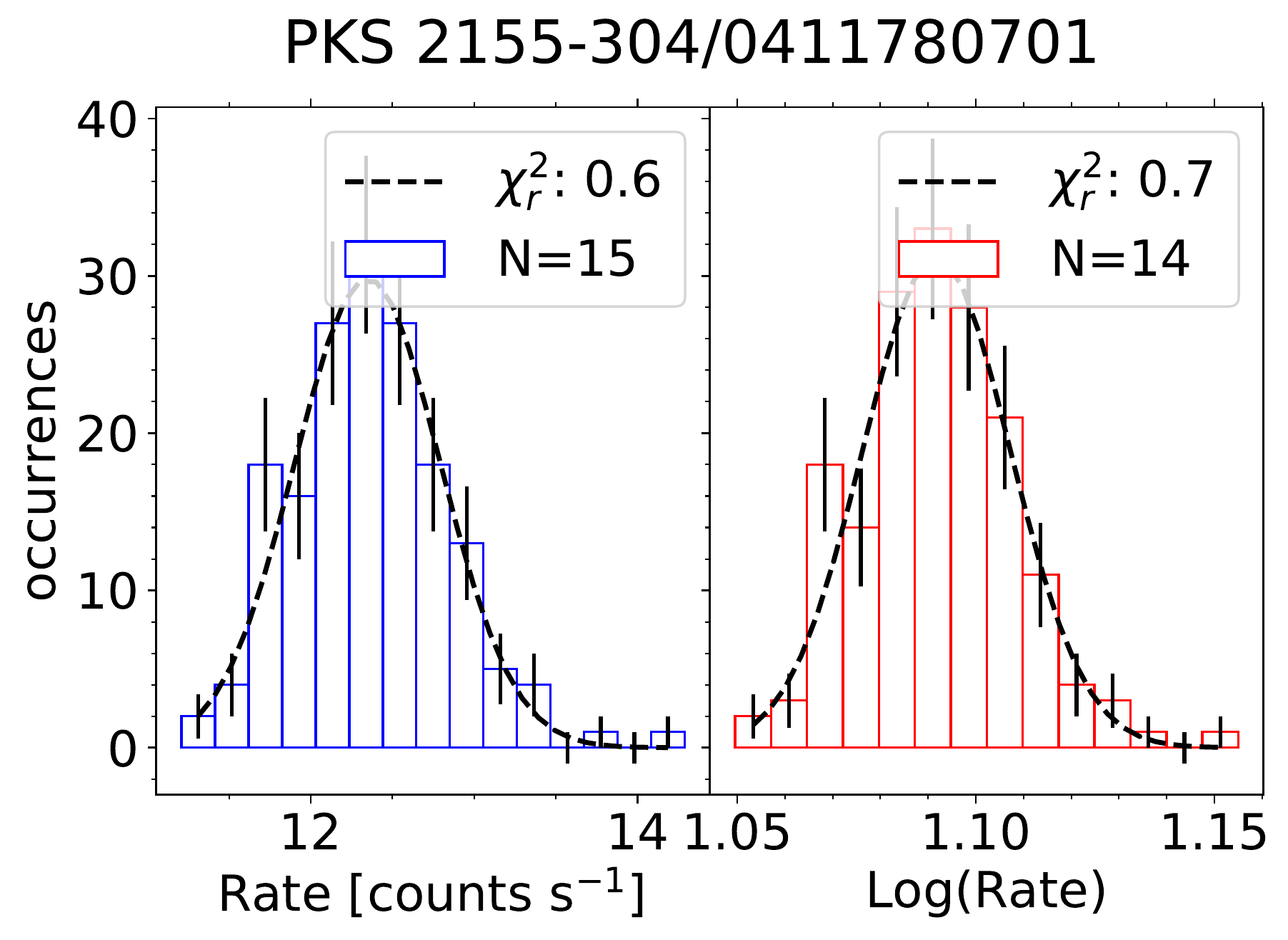}
\includegraphics[scale=0.275]{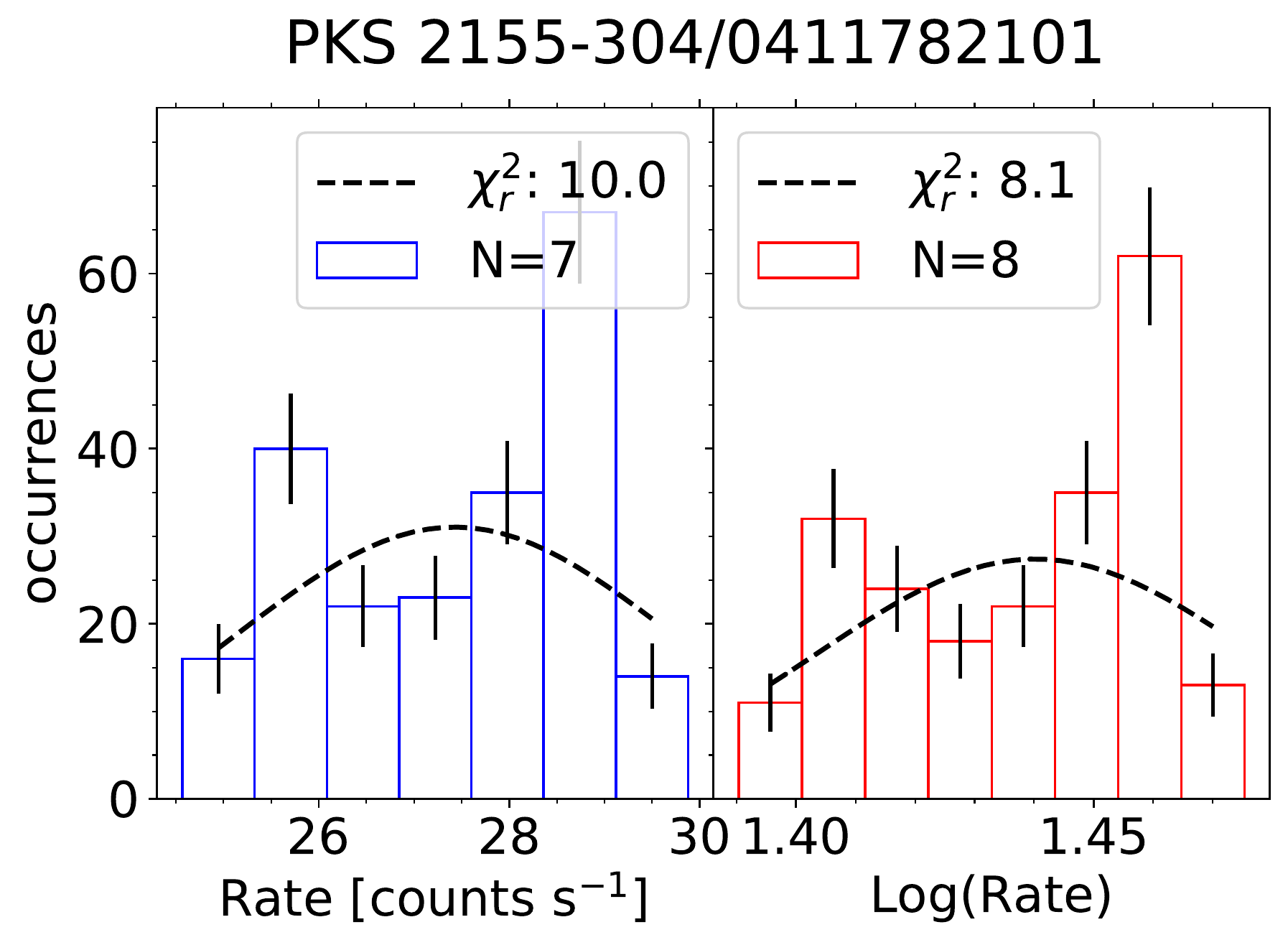}
\includegraphics[scale=0.275]{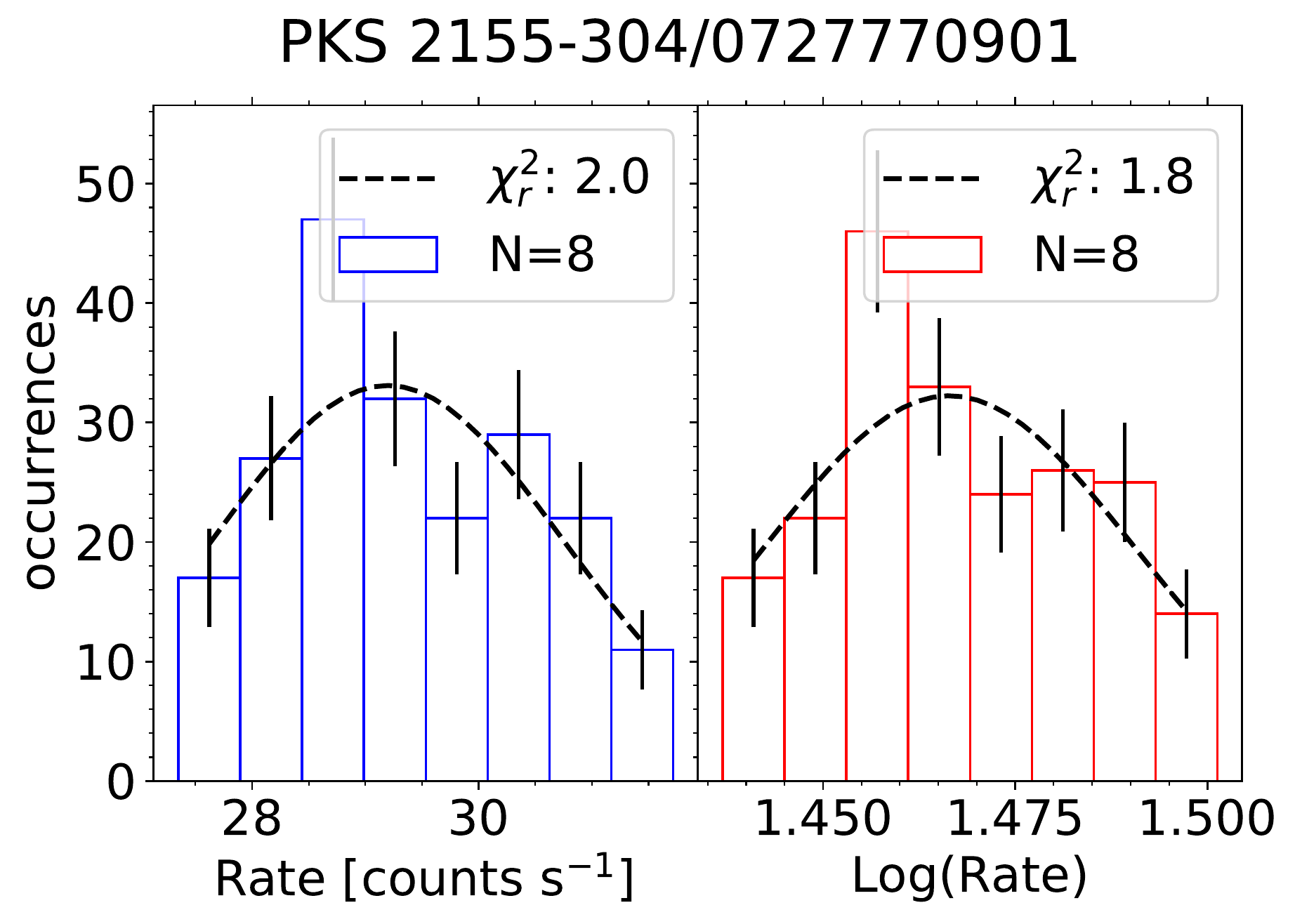}
\includegraphics[scale=0.275]{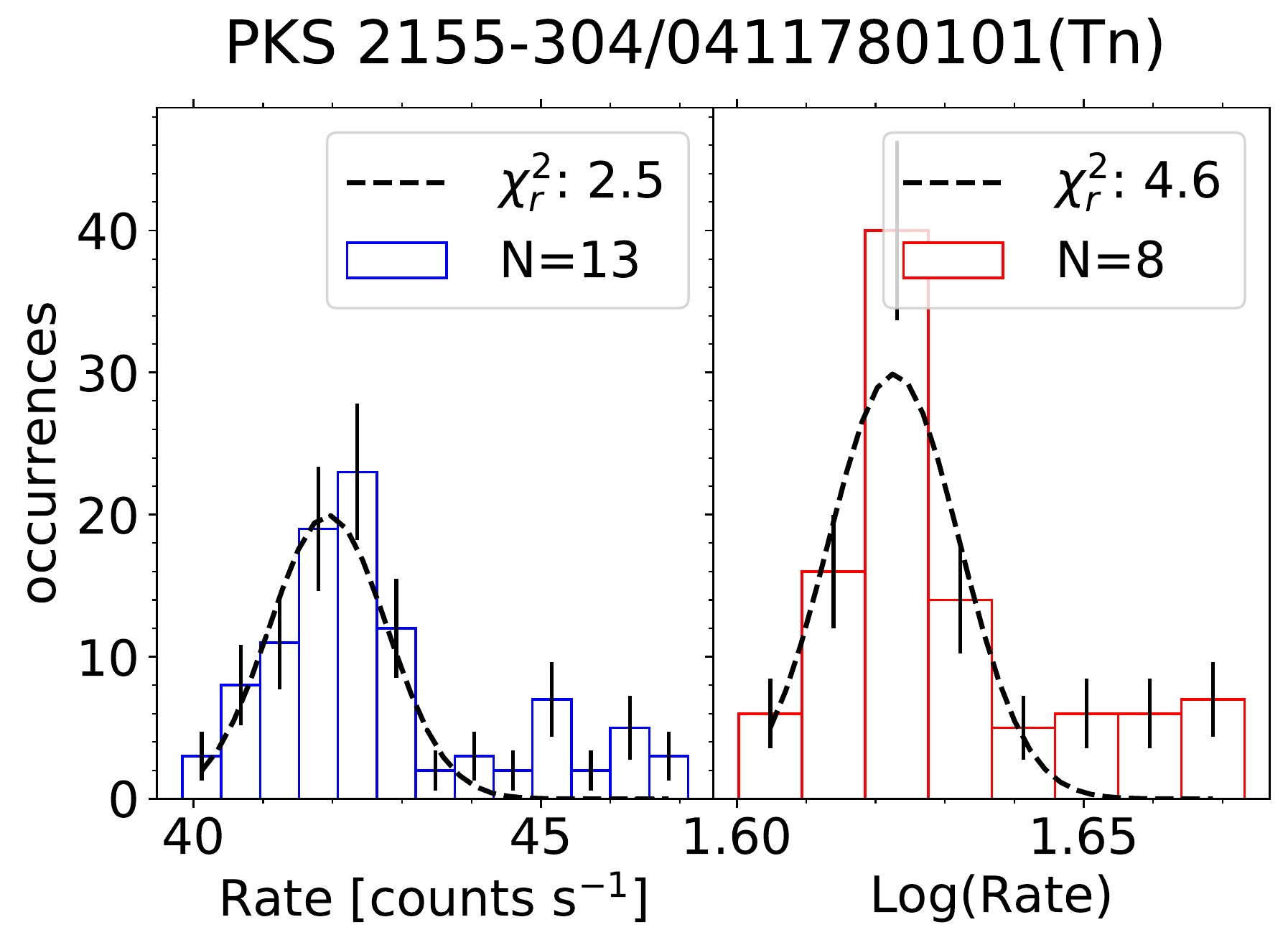}
\includegraphics[scale=0.275]{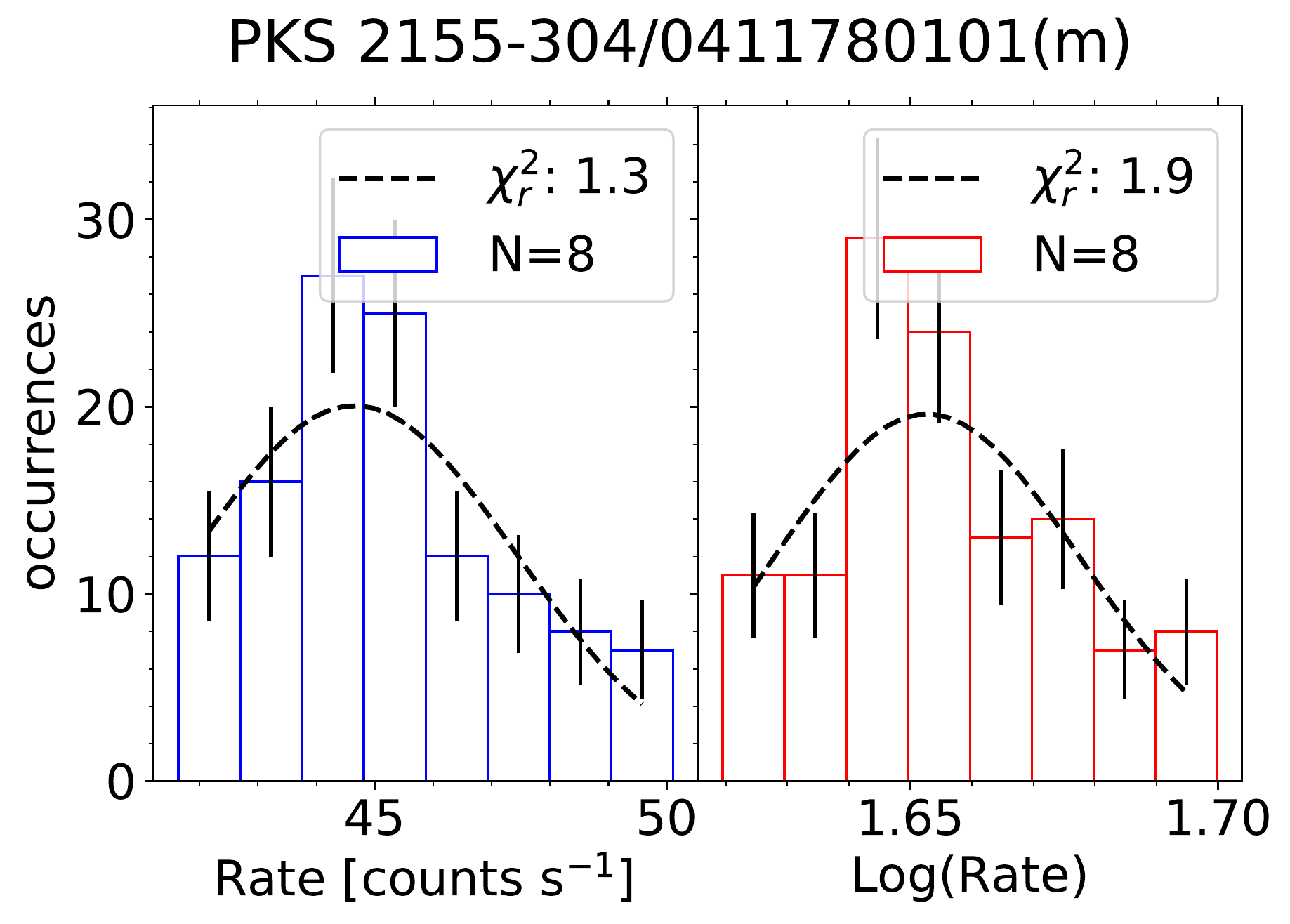}
\includegraphics[scale=0.275]{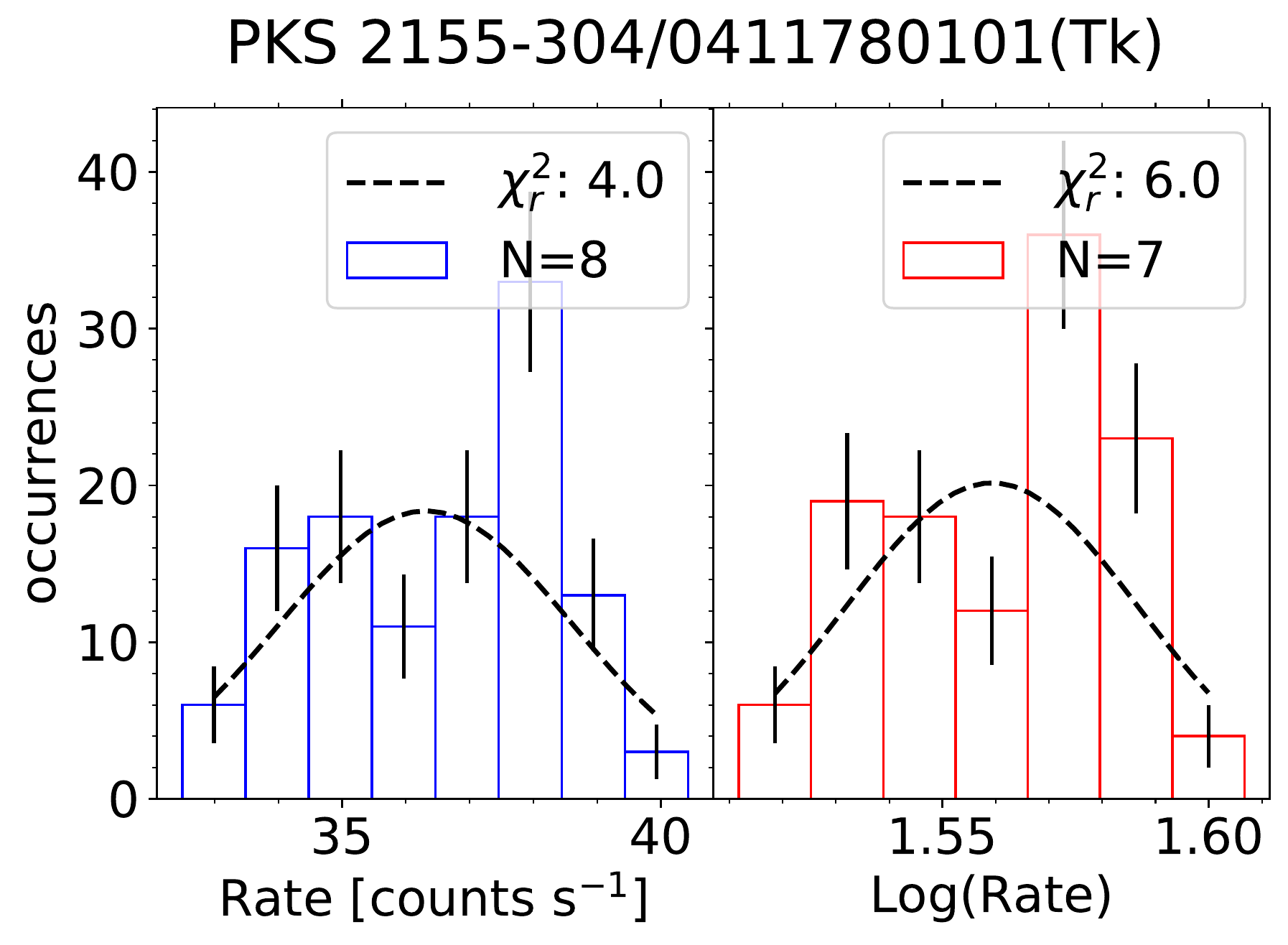}
\caption{Histograms of total count rate (blue bars) and the logarithm (log)
of the total count rate (red bars) for each ID. The dashed curve in the respective
window is a Gaussian fit to the histogram and \(\chi^2_r\) is the best-fit 
reduced-\(\chi^2\) value. N denotes the number of bins.}
\label{fig:hist}
\end{figure}

A histogram is one of the ways to visualize variations, its extent, and explore any
trend/scales as well as understand the nature of variability regardless of time.
Motivated by theoretical considerations and  observational finding of blazars
flux variation
being log-normal \citep[e.g.][]{2016ApJ...822L..13K,2017ApJ...849..138K}, we derived two different
histograms -- one from the 0.3 -- 10 keV count rate and the other from the logarithm\footnote{unless stated otherwise, logarithm (log) throughout this work means log
to the base 10} of this count rate using the Knuth method \citep{KNUTH2019102581}.
Figure \ref{fig:hist} shows the two histograms for each ID and a Normal fit to these
histograms along with the reduced-\(\chi^2\) (\(\chi^2_r\)) values.

Interestingly, if we choose the lowest \(\chi^2_r\) as one representing the data better,
a Normal fit is better compared to a log-Normal for both the sources except for the
three IDs -- 0411780501, 0411782101, and 0727770901 of PKS 2155-304. For 0411780501,
the histogram rather appears more consistent with a constant while for 0727770901,
log-Normal is marginally better. For 0124930301 and 0411780401, even though Normal
is favored, the inferences are unphysical as the peak rate returned by the fit is
way higher than anything reported in the literature so far.

We, additionally, also performed the Anderson-Darling (AD) test to see which of the
two histograms: Normal or log-Normal is a better representation of the data. The
result of the AD test is listed in the last column of Table \ref{tab:test}. Interestingly, the AD test always favored a log-Normal over Normal whether or not
the test statistic is within the critical significance level values. Further,
the test values for log-Normal was consistently lower than the test values for
a Normal. The '-' in Table
\ref{tab:test} means that the test result was greater than the critical value corresponding to p=0.01. These inferences are contrary to the one inferred 
from \(\chi^2_r\) fitting.

\subsection{RMS-Flux relation}
Photon counting being Poissonion in nature, has inherent variation. Whether the
observed variability is due to this statistical fluke or real -- intrinsic to
the source is normally estimated by measuring variance. If the variance of the 
data is larger than the mean-error-squared, then the source is variable and
the variability could be intrinsic to the source. 

\begin{figure}[H]
\centering
\includegraphics[scale=0.27]{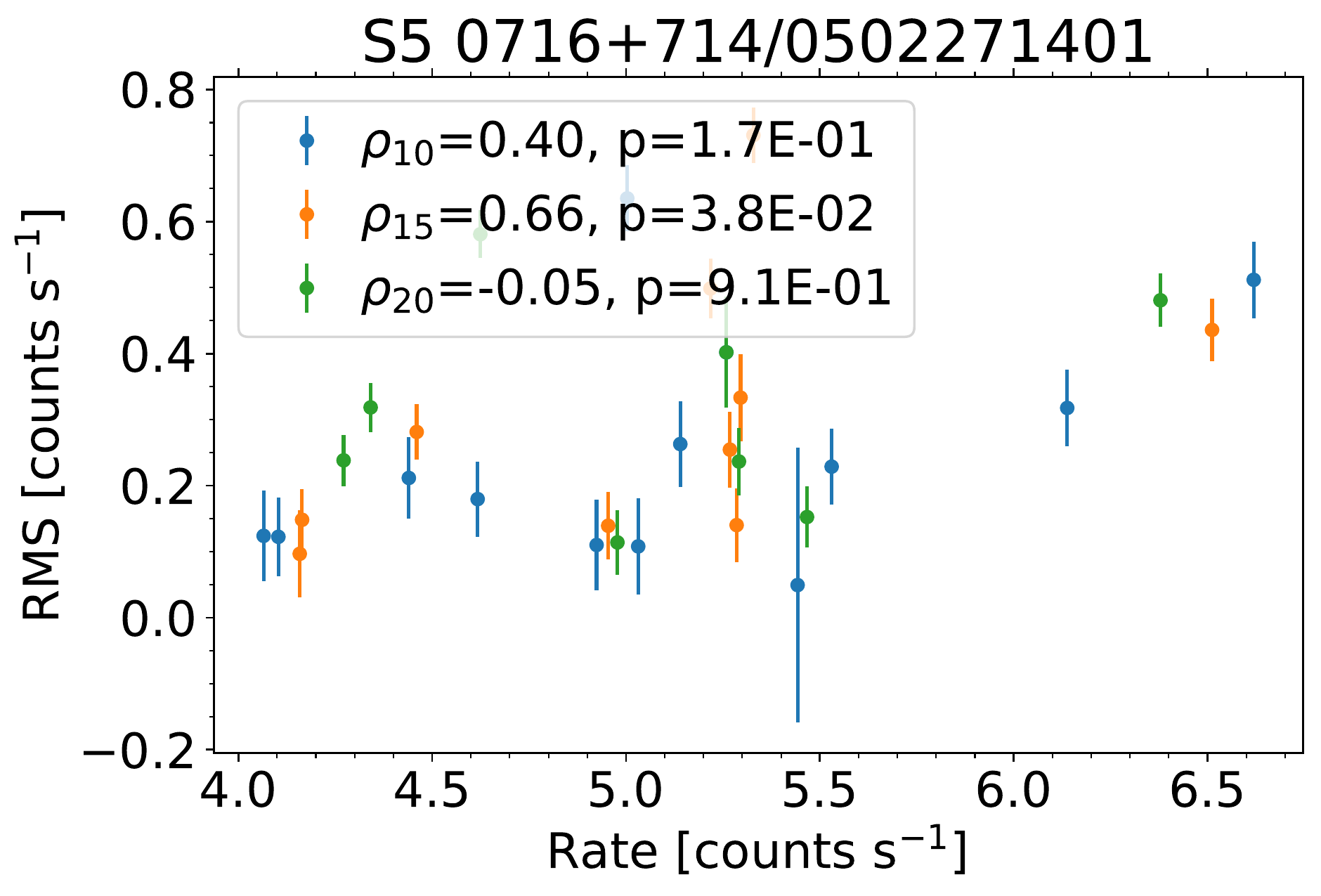}
\includegraphics[scale=0.27]{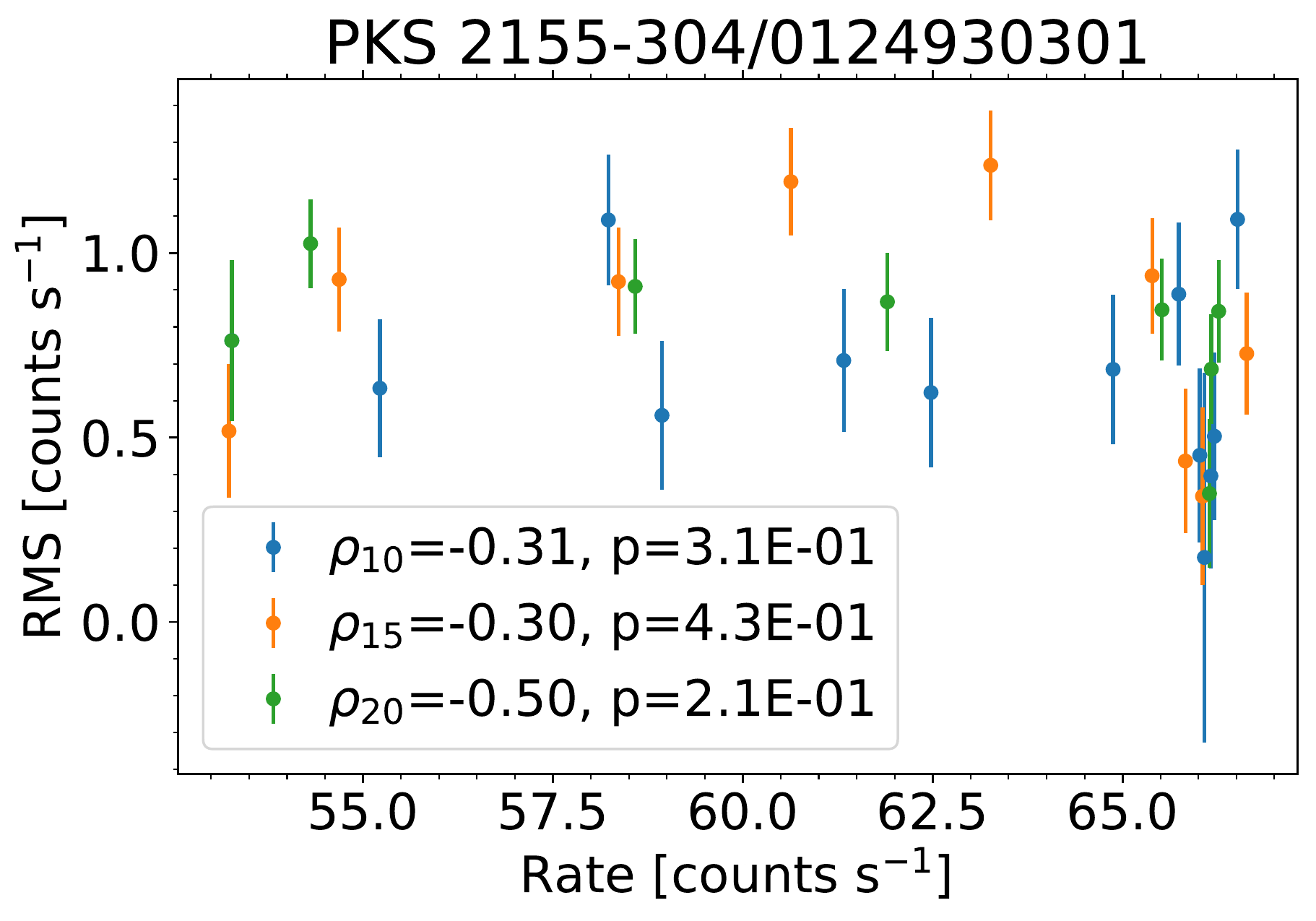}
\includegraphics[scale=0.27]{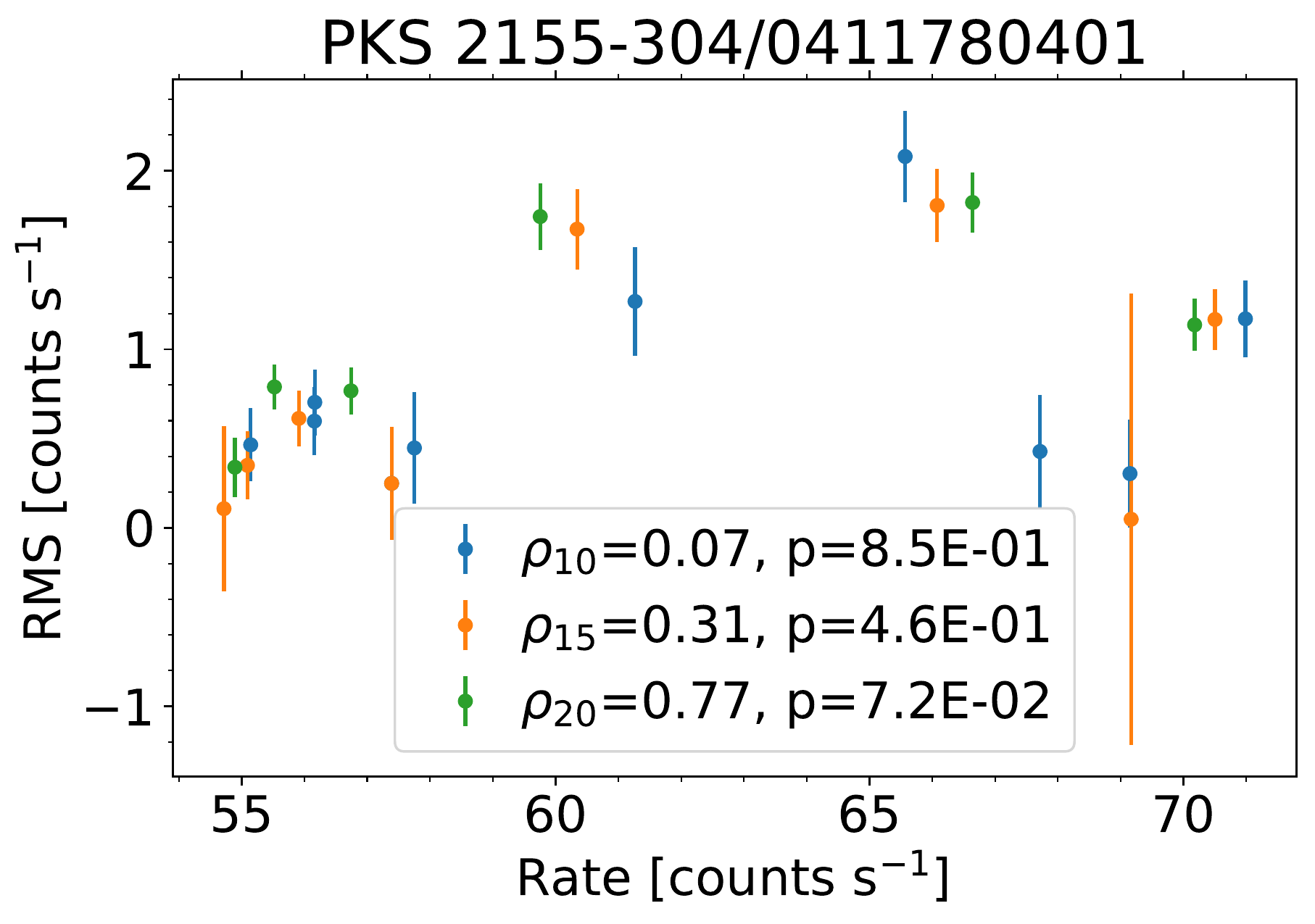}
\includegraphics[scale=0.27]{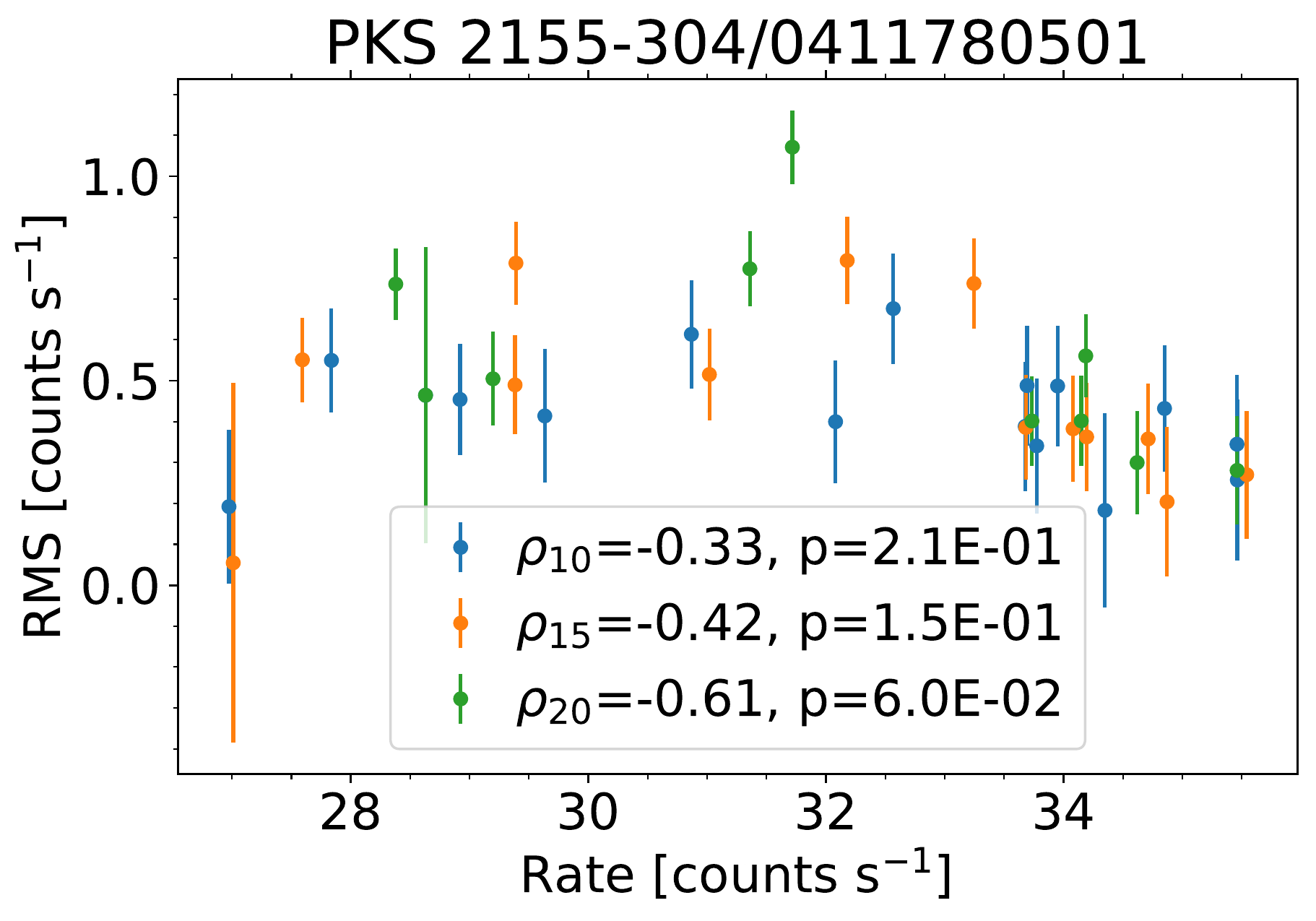}
\includegraphics[scale=0.27]{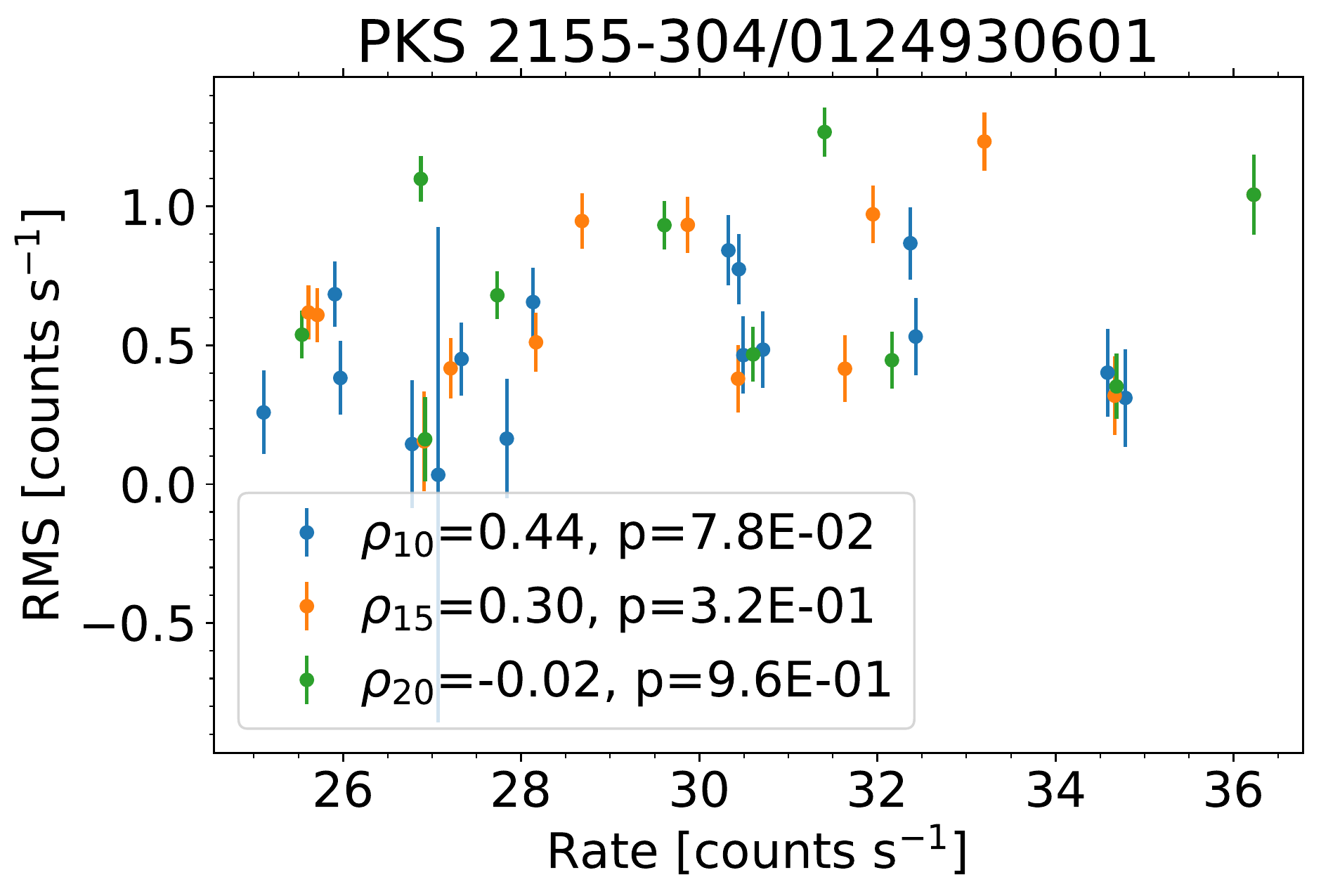}
\includegraphics[scale=0.275]{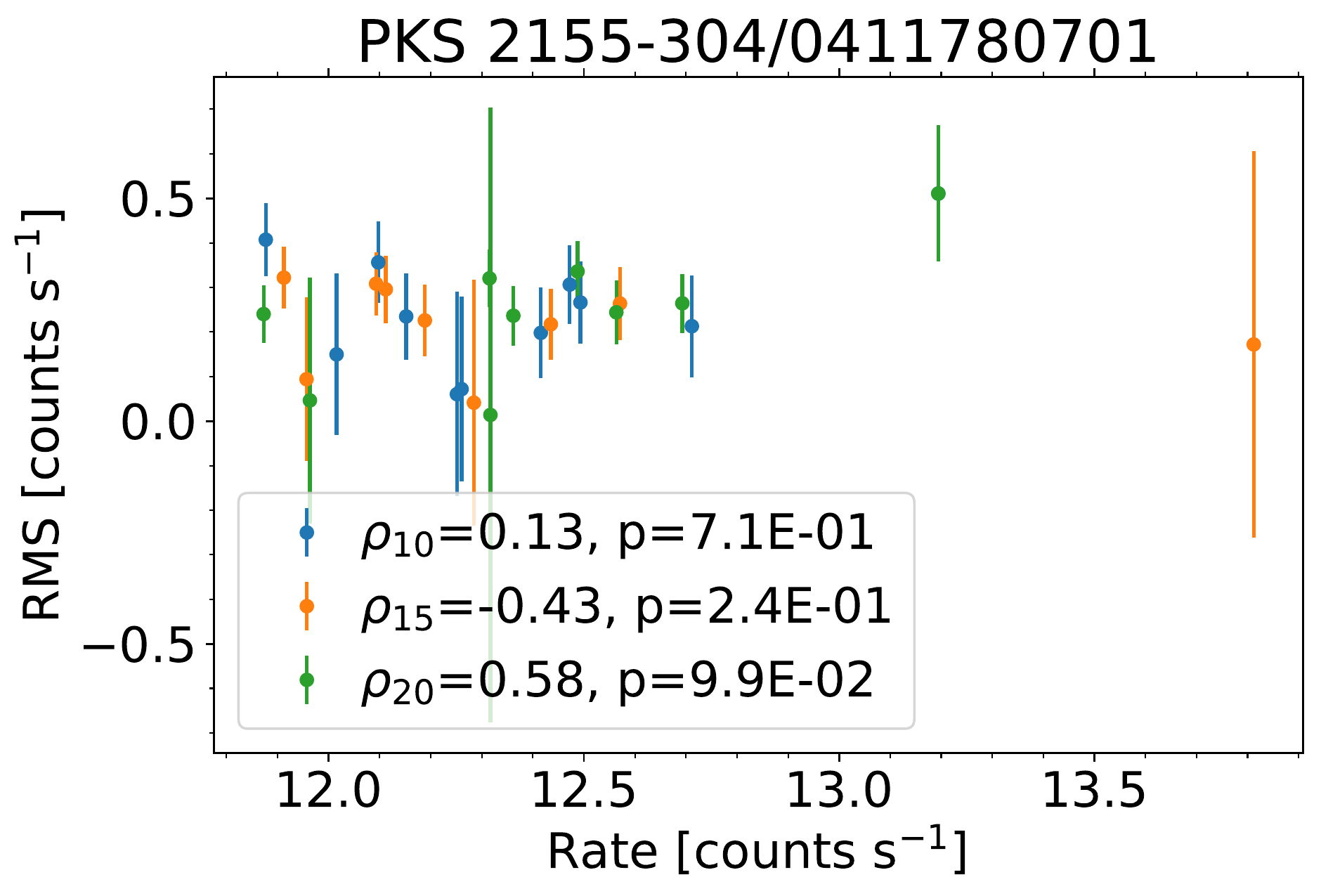}
\includegraphics[scale=0.275]{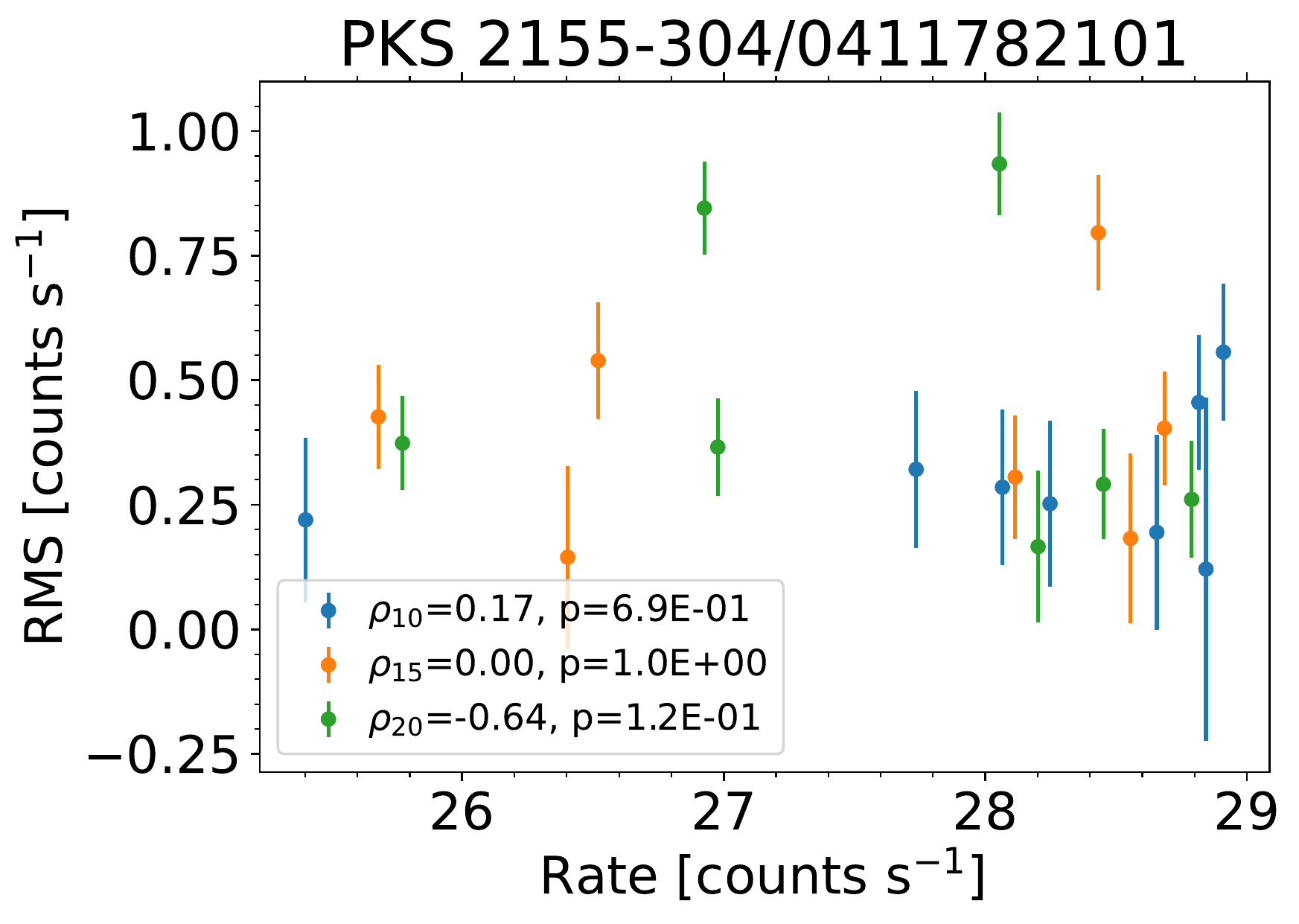}
\includegraphics[scale=0.275]{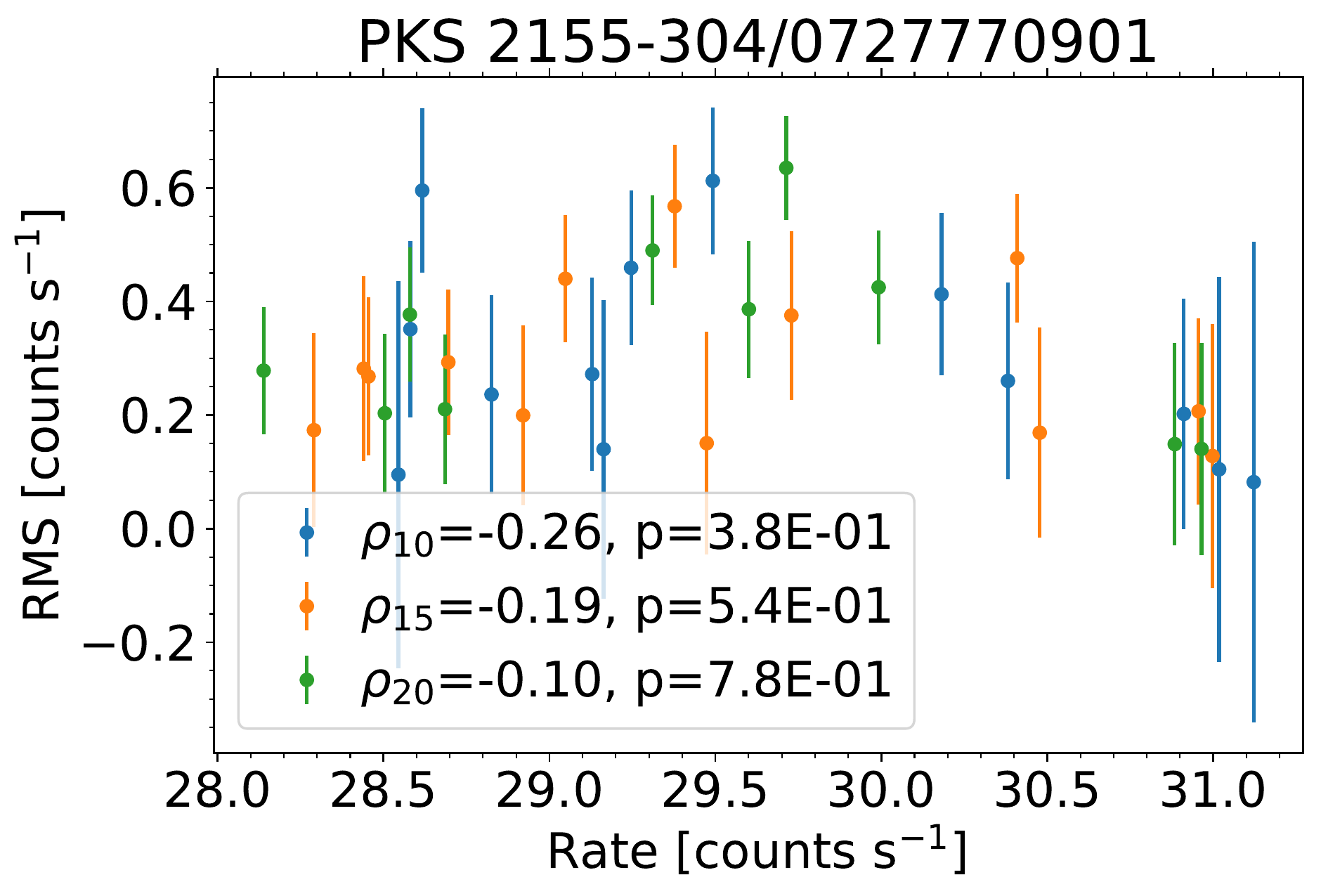}
\includegraphics[scale=0.275]{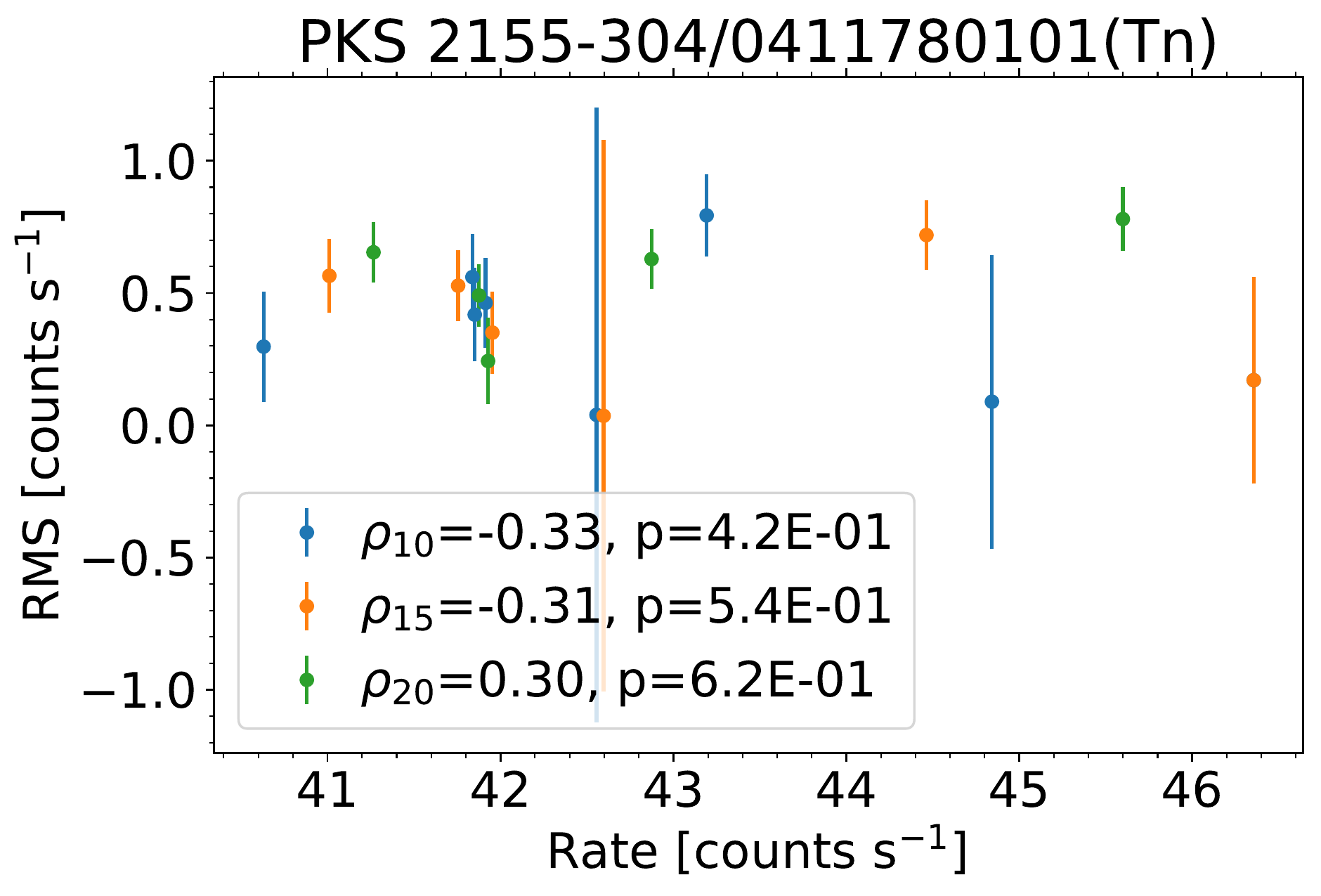}
\includegraphics[scale=0.275]{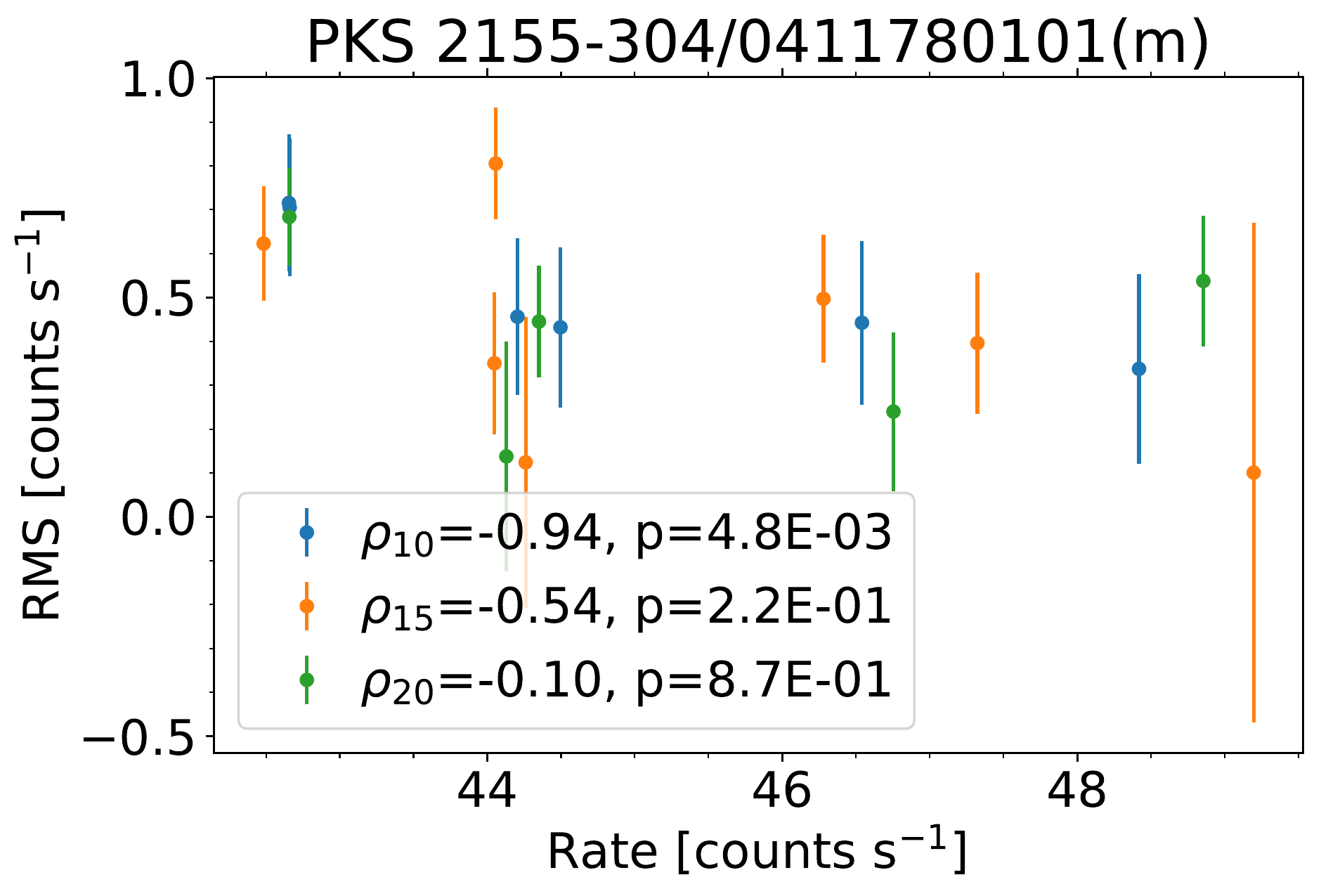}
\includegraphics[scale=0.275]{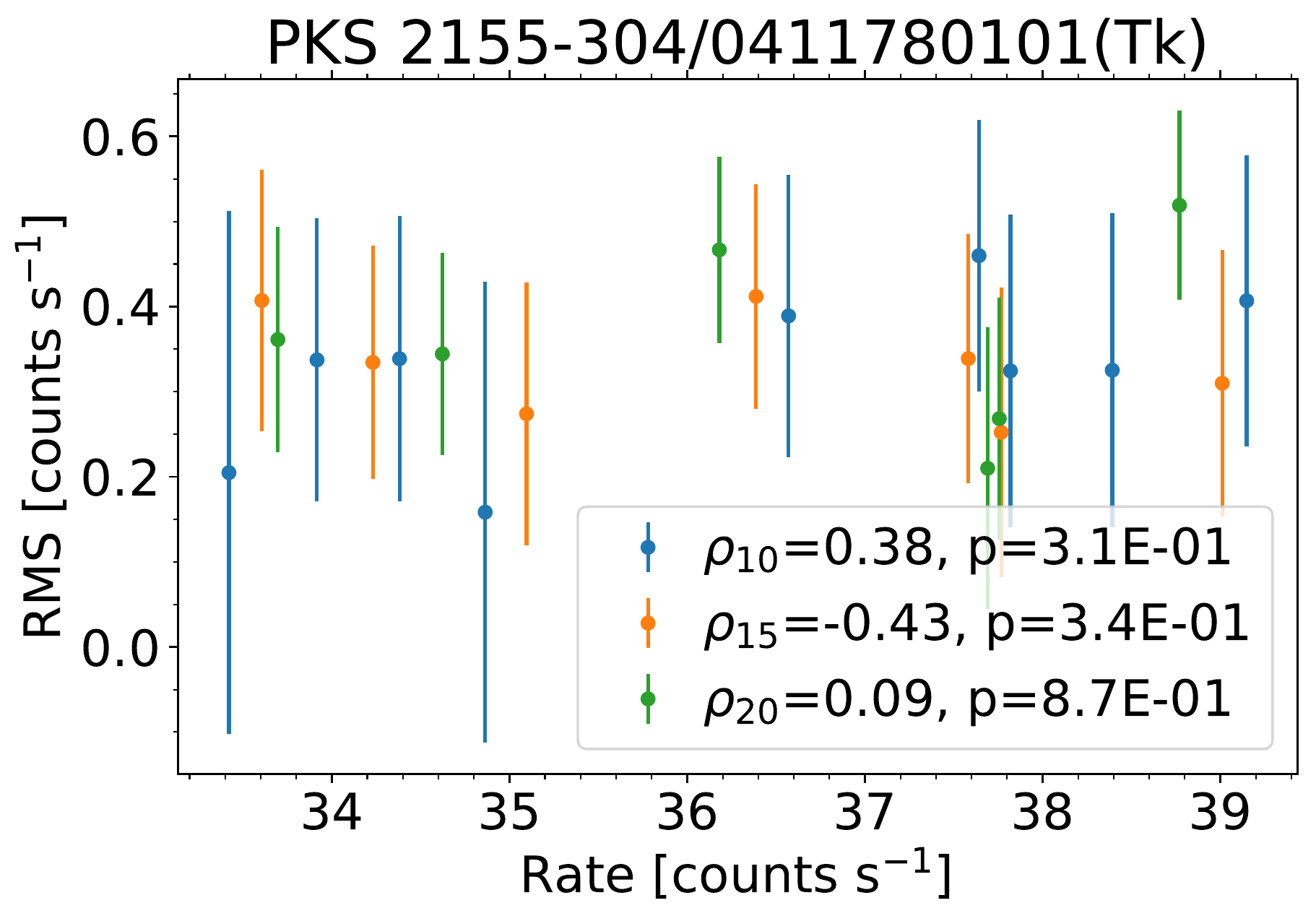}
\caption{RMS-flux relation using 10, 15, and 20 data points per bin for each ID 
along with the corresponding Spearman correlation coefficient ($\rho$) and p-value.}
\label{fig:rmsF}
\end{figure}

An important characteristic of accretion-powered sources and even blazars has been
a linear relationship between flux/count rate and RMS (root-mean-squared) variation
\citep{2003MNRAS.345.1271V,2005ApJ...629..686Z,2017ApJ...849..138K}. However, recent observations
have reported a few contrary results as well \citep[and reference therein]{2020ApJ...895...90S}. Figure \ref{fig:rmsF} shows the RMS-flux relation for the current
sample using 10, 15, and 20 data points per bin along with the corresponding Spearman correlation
coefficient and p-value (p$<$0.01 means significant correlation or anti-correlation
depending on the sign of $\rho$). We also verified the RMS-flux relation using more data
points per bin (e.g. 25, 30) and the outcome remains similar. None of these show
any significant correlation between RMS and count rate. It should be noted that
during the calculation, a few of the segments had negative variances due to measurement
error dominating the variations. We simply rejected variance in these cases.

\section{Discussion and Conclusion}\label{sec:Discuss}
In this work, we explored the statistical properties of short-term (300s binned)
X-ray variability of blazar S5 0716+714 and PKS 2155-304 during high variability
phases using long-exposure {\it XMM-Newton} EPIC-PN camera observations. The only
ID of S5 7016+714 shows about a factor of two change in the count rate while
 the rest of the IDs belongs to PKS 2155-304
and represents different intensity and variability states of the source. For the later,
the count rate changes by a factor of 1.2 -- 1.5 within each ID but the intensity
states covered by these IDs correspond up to a factor of 6.5. Test of stationarity
using ADF and KPSS shows that the variability is highly non-stationary both in normal
and logarithmic space (ref. Table \ref{tab:test}) except probably for S5 0716+714
where the KPSS test suggests trend-stationarity. In fact, short-term X-ray variability 
of blazars has been normally found to be non-stationarity \citep[e.g.][]{2005A&A...443..397B,2005ApJ...629..686Z,2020ApJ...897...25B}.
This suggests that using Fourier methods to investigate variability can result in
erroneous inferences. However, the difference light curve derived by subtracting
the previous value from the next makes all the light curves stationary except IDs
0411780501 and 0411780601 of PKS 2155-304, indicating the auto-regressive 
integrated moving average (ARIMA) method for timing studies as more appropriate
\citep[e.g.][]{2020ApJ...897...25B}.

In high states of the source, the observations show a complex temporal evolution of
intensity vis-a-vis HR (ref Fig. \ref{fig:lc}). S5 0716+714 shows anti-correlation
between count rate evolution with HR while for PKS 2155-304, the behavior is more
complex and involved. At low to moderate count rates ($\lesssim 50$), count rate correlates with HR except for the thin filter observation in 0411780101 where towards
the end, HR anti-correlates with count rate.
For the brightest two, the evolution is complex with correlation and anti-correlation
both, indicating spectral evolution over time scales of a few 10s of ks. Such strong
X-ray spectral evolution has been seen over similar and much smaller time scales 
in Mrk 401 \citep[e.g.][]{2005A&A...443..397B} and indicates that X-ray SED derived
using full observation is not a true representative of the spectral state of the source
but rather an average one \citep{2017ApJ...850..209G}. In general, anti-correlation
indicates a steepening of the X-ray spectrum, suggesting radiative cooling dominated evolution while correlation
suggests activity driven by injection or particle acceleration \citep[e.g.][]
{2014ApJ...796...61K}. However, the latter can also be due to additional emission
regions contributing to the emission as has been argued in the case of spectral changes
in Mrk 421 \citep[e.g.][]{2005A&A...443..397B}.

An interesting statistical property of AGNs X-ray variability has been a log-normal
flux distribution and a linear relation between variance and flux. Similar behavior
has been reported for blazars emission as well, both on long- and short-term observations
\citep[e.g.][]{2009A&A...503..797G,2010A&A...520A..83H,2016ApJ...822L..13K,
2017ApJ...849..138K}. Here, except for the one ID (0411780701) of PKS 2155-304, none
of the histograms are consistent with either normal or lognormal distribution based
on the \(\chi^2\) method. In fact, the histograms show diverse profiles with normal
distribution returning systematically lower \(\chi^2\) statistic value than a log-normal,
indicating a preference for normal. On the contrary, the AD test (ref table \ref{tab:test}) always favors a log-normal with test-statistic lower than that for
a normal, regardless of whether the outcome is significant or not. Based on the AD
test, three observations
were found to be consistent with lognormal or both (p-value $<0.01$). Similarly,
none of the observations show any significant linear correlation between the count
rate and excess variance (ref fig. \ref{fig:rmsF}) except the case of medium filter
observation of ID 0411780101 with 10 data points per bin. In short, data considered
here neither show log-normal intensity distribution nor a linear correlation between excess variance and intensity. 

Based on simulations and the fact that blazars jet are highly magnetized system, magnetic
reconnection has been favored as a potential mechanism for rapid and luminous flares 
\citep[e.g.][]{2013MNRAS.431..355G}. One of the statistical models based on magnetic
reconnection called minijets-in-a-jet \citep{2012A&A...548A.123B} predicts diverse
histogram profiles, ranging from  a power-law to one consistent with a log-normal 
depending on the number of emission region contributing to the emission. However,
the RMS-flux relation is always linear. Our results here disfavor such a model.
Nonetheless, the difference in long and short term statistical behavior is very
different and intriguing. Whether incorporating cooling and/or acceleration/injection
can result in such changes remains to be seen.


\vspace{6pt} 




\funding{PK acknowledge funding from ARIES Aryabhatta Postdoctoral fellowship
(AO/A-PDF/770). MP thanks the
financial support of UGC, India program through DSKPDF fellowship (grant No.
BSR/2017-2018/PH/0111).}

\acknowledgments{This research has made use of archival data of XMM-Newton observatory,
an ESA science mission directly funded by ESA Member States and NASA by the NASA
Goddard Space Flight Center (GSFC).}

\conflictsofinterest{Declare conflicts of interest or state ``The authors declare no conflict of interest.'' Authors must identify and declare any personal circumstances or interest that may be perceived as inappropriately influencing the representation or interpretation of reported research results. Any role of the funders in the design of the study; in the collection, analyses or interpretation of data; in the writing of the manuscript, or in the decision to publish the results must be declared in this section. If there is no role, please state ``The funders had no role in the design of the study; in the collection, analyses, or interpretation of data; in the writing of the manuscript, or in the decision to publish the results''.} 

\abbreviations{The following abbreviations are used in this manuscript:\\

\noindent 
\begin{tabular}{@{}ll}
KPSS & Kwiatkowski-Phillips-Schmidt-Shin \\
ADF & Augumented Dickey-Fuller \\
AGNs & Active Galactic Nuclei\\
IDV & Intra-day variability\\
LTV & Long term variability \\
STV & Short-term variability\\
AD & Anderson-Darling \\
HR & Hardness ratio
\end{tabular}}


\reftitle{References}


\externalbibliography{yes}
\bibliography{galaxies_X}




\end{document}